\documentclass[tightenlines,floats,aps,nofootinbib,prd,onecolumn,preprintnumbers]{revtex4-2}

\usepackage[dvipdfmx]{graphicx}
\usepackage{amsmath,amssymb,algorithm,algorithmic,bm,color,mathrsfs,subcaption,tikz,multirow}
\usepackage[pdftex,colorlinks=true,linkcolor=blue,citecolor=cyan]{hyperref}

\usetikzlibrary{positioning}

\allowdisplaybreaks[1]

\newcommand{\xx}{\boldsymbol{x}}

\newcommand{\Mpl}{M_{\rm pl}}

\begin{document}

\title{CMB constraints on DHOST theories}

\author{Takashi Hiramatsu$^1$\footnote{Email: hiramatz@rikkyo.ac.jp}}
\affiliation{
$^1$ Department of Physics, Rikkyo University, Toshima, Tokyo 171-8501, Japan
}

\begin{abstract}
We put constraints on the degenerate higher-order scalar-tensor (DHOST) theories using the Planck 2018 likelihoods.
In our previous paper, we developed a Boltzmann solver incorporating the effective field theory parameterised by the six time-dependent functions,
$\alpha_i$ $(i={\rm B},{\rm K},{\rm T},{\rm M},{\rm H})$ and $\beta_1$, which can describe the DHOST theories. 
Using the Markov-Chain Monte-Carlo method with our Boltzmann solver,
we find the viable parameter region of the model parameters characterising the DHOST theories and the other standard cosmological parameters.
First, we consider a simple model with $\alpha_{\rm K} = \Omega_{\rm DE}(t)/\Omega_{\rm DE}(t_0)$, $\alpha_{\rm B}=\alpha_{\rm T}=\alpha_{\rm M}=\alpha_{\rm H}=0$ and $\beta_1=\beta_{1,0}\Omega_{\rm DE}(t)/\Omega_{\rm DE}(t_0)$ in the $\Lambda$CDM background where $t_0$ is the present time and obtain $\beta_{1,0}=0.032_{-0.016}^{+0.013}$ (68\% c.l.).
Next, we focus on another theory given by $\mathcal{L}_{\rm DHOST} = X + c_3X\Box\phi/\Lambda^3+ (\Mpl^2/2+c_4X^2/\Lambda^6)R + 48c_4^2X^2/(M_{\rm pl}^2\Lambda^{12}+2c_4\Lambda^6X^2)\phi^\mu\phi_{\mu\rho}\phi^{\rho\nu}\phi_\nu$
 with $X:=\partial_\mu\phi\partial^{\mu}\phi$ and two positive constant parameters, $c_3$ and $c_4$.
In this model, we consistently treat the background and the perturbations, and obtain $c_3 = 1.59^{+0.26}_{-0.28}$ and the upper bound on $c_4$, $c_4<0.0088$ (68\% c.l.).
\end{abstract}

\preprint{RUP-22-11}

\maketitle

\section{Introduction}
\label{sec:intro}

Extended theories of gravity beyond general relativity (GR) have been extensively discussed to test gravity in various phenomenological aspects.
One can extend GR by adding another scalar degree of freedom
mediating gravity, called scalar-tensor theories. 
Such an extended theory should evade pathological properties arising from the scalar degree of freedom, such as the ghost instability, and thus the equation-of-motion should be described by second-order differential equations in time.
The Horndeski theory is known as a healthy theory and has been extensively investigated\cite{Horndeski:1974wa,Deffayet:2011gz,Kobayashi:2011nu} (for review, see \cite{Kobayashi:2019hrl}), and its extension, known as the Gleyzes-Langlois-Piazza-Vernizzi (GLPV) theory or beyond Horndeski theory, has also been well studied \cite{Gleyzes:2014dya,Gleyzes:2014qga}. These theories contain many well-known scalar-tensor theories as subclasses, for instance, the k-essence model, $f(R)$ gravity and the Galileon theories. The degenerate higher-order scalar-tensor (DHOST) theory is known as the most general theory of the single-scalar-tensor theories, including the Horndeski theory and the GLPV theory \cite{Langlois:2015cwa,Langlois:2015skt} (for review, see \cite{Langlois:2018dxi}). In particular, we focus on the Type-I quadratic DHOST theories (categorised to class Ia \cite{BenAchour:2016cay} and class N-I \cite{Crisostomi:2016czh}) characterised by eight arbitrary functions of the scalar field $\phi$ and $X:=\partial_\mu\phi\partial^\mu\phi$, which is phenomenologically interesting in the astrophysical and cosmological contexts. 
As the action of the DHOST theories contains the higher-derivatives, the equation-of-motion has fourth-order temporal derivatives, and thus the extra two scalar degrees of freedom exist. One of them is a ghost degree of freedom, known as the Ostrogradsky ghost. One can remove the ghost degree of freedom and obtain the second-order equations if the kinetic matrix of the two scalar degrees of freedom is degenerate. The condition for the degeneracy is described as the relations among the arbitrary functions in the action.

Many groups have investigated the phenomenological aspects of the extended theories up to the GLPV theory, providing many ways to test the theories from the observations of CMB, the large-scale structure, and the gravitational waves \cite{DeFelice:2015isa,DAmico:2016ntq,Renk:2016olm,Kreisch:2017uet,Namikawa:2018erh,Noller:2018wyv,Peirone:2019yjs,Traykova:2019oyx}.
Besides the cosmological-scale observations, the existence of compact stars like neutron stars can put constraints on them through the partial breaking of the screening mechanism \cite{Sakstein:2015zoa,Sakstein:2015aac,Jain:2015edg,Sakstein:2016ggl,Sakstein:2016lyj,Salzano:2017qac,Babichev:2016jom,Sakstein:2016oel}.

In this paper, we study the angular power spectrum of the CMB anisotropies in the DHOST theories beyond the GLPV theory.
We have developed a Boltzmann solver incorporating the DHOST theories \cite{Hiramatsu:2020fcd}.
Using the numerical code, we have demonstrated the qualitative impact of the model parameters on the angular power spectra of the CMB anisotropies using the Fisher information matrix. 
The Fisher analysis assumes a Gaussian likelihood, and we have computed the 1-sigma uncertainties with fixed cosmological parameters such as the amplitude of the primordial perturbations and the Hubble parameter. 
We improve this study using the Markov-Chain Monte-Carlo (MCMC) method and compute the confidence intervals of the model parameters using the Planck 2018 likelihoods \cite{Planck:2018vyg}.
Note that there are pioneering works to put constraints on the model parameters in the DHOST theories focusing on compact stars \cite{Kobayashi:2018xvr,Chagoya:2018lmv}. Our present study is based on cosmological observations, providing a new test for the theories.

To describe the DHOST theories, we employ the effective field theory (EFT) approach \cite{Gubitosi:2012hu,Bloomfield:2012ff,Gleyzes:2013ooa,Bloomfield:2013efa,Piazza:2013coa} characterised by six time-dependent functions, $\alpha_i(t)$ with $i={\rm K},{\rm B},{\rm T},{\rm M},{\rm H}$ and $\beta_1(t)$ \cite{Gleyzes:2014rba,Gleyzes:2015pma,DAmico:2016ntq,Langlois:2017mxy}. As mentioned before, the DHOST theories cover a huge number of the single-scalar-tensor theories by choosing the arbitrary functions in the action. The EFT parameters can be written in terms of them. The Horndeski theory and its subclass theories can be described by the first four functions, $\alpha_i$ with $i={\rm K},{\rm B},{\rm T},{\rm M}$. The remaining parameters, $\alpha_{\rm H}$ and $\beta_1$, are called beyond-Horndeski parameters. While the GLPV theory requires $\alpha_{\rm H}\ne 0$ and $\beta_1=0$, the DHOST theories correspond to the theories with $\beta_1\ne 0$. The advantages of employing the EFT approach are that we can explore the theories without specifying the arbitrary functions of $\phi$ and $X$ in the original action and that we can treat the background sector, i.e. cosmic expansion history, separately from the perturbative sector. This approach is convenient to study comprehensively how significant the parameters characterising the DHOST theories affect the CMB angular power spectrum. We consider a subclass theory of the DHOST theories where $\alpha_{\rm H}=\alpha_{\rm M}=\alpha_{\rm T}=\alpha_{\rm B}=0$ and compute the confidence intervals of the parameter $\beta_1$.

Next, we change the approach and treat the background and perturbative sectors consistently by specifying the arbitrary functions contained in the DHOST theories. In particular, we adopt the parameterisation proposed by Crisostomi and Koyama (CK) \cite{Crisostomi:2017pjs}. With this treatment, we first solve the background equations for the scalar field and the scale factor and solve the perturbation equations in this background. Then we estimate the confidence intervals of the model parameters.

Our Boltzmann solver developed in our previous paper \cite{Hiramatsu:2020fcd} incorporates the EFT equations with the EFT parameters up to $\beta_1$. In the EFT approach, we assume that the six EFT parameters are proportional to $\Omega_{\rm DE}(t)$, the fractional amount of the dark energy, following the same setup as in Ref.~\cite{DAmico:2016ntq}.
In the CK model, the EFT parameters are given by functions of the background scalar field $\phi$ and $X$ that are determined by solving the background equation for $\phi$. Hence the model parameters characterising this model affect both the background and the perturbations. We use the MCMC method incorporating the Metropolis algorithm to 
compute the confidence intervals of the model parameters in the EFT approach and the CK model.

This paper is organised as follows. In Sec.~\ref{sec:DHOST}, we briefly review the DHOST theories and their effective description. In Sec.~\ref{sec:model}, we explain two concrete theoretical models and briefly mention the numerical setup. Then we show the results of the MCMC simulations in Sec.~\ref{sec:results} and give a conclusion in Sec.~\ref{sec:conclusion}.

Throughout this paper, we use the unit with $c=\hbar=1$.

\section{Type-I quadratic DHOST theory}
\label{sec:DHOST}

\subsection{Background}

We consider the Type-I quadratic DHOST theory \cite{Langlois:2015cwa,Langlois:2015skt} with a minimally-coupled matter, whose action is given as
%
\begin{align}
 S &= \int\!d^4x\,\sqrt{-g}\,\mathcal{L}_{\rm DHOST} + \int\!d^4x\,\sqrt{-g}\,\mathcal{L}_{\rm m},
\label{eq:def_action}
\end{align}
%
with
%
\begin{align}
 \mathcal{L}_{\rm DHOST} &:= P(\phi,X) + Q(\phi,X)\Box\phi + f_2(\phi,X){}^{(4)}R
    + \sum_{i=1}^{5}a_{i}(\phi,X)\mathcal{L}_{i}
, \label{eq:DHOST_action}
\end{align}
%
where $P,Q,f_2$ and $a_{i}$ are functions of $\phi$ and $X:=\partial_\mu\phi\partial^\mu\phi$ and
%
\begin{align}
 \mathcal{L}_1 &:= \phi_{\mu\nu}\phi^{\mu\nu}, \;
 \mathcal{L}_2 := (\Box\phi)^2, \;
 \mathcal{L}_3 := (\Box\phi)\phi^\mu\phi_{\mu\nu}\phi^\nu,
\notag \\ 
 \mathcal{L}_4 &:= \phi^\mu\phi_{\mu\rho}\phi^{\rho\nu}\phi_\nu, \;
 \mathcal{L}_5 := (\phi^\mu\phi_{\mu\nu}\phi^\nu)^2,
\end{align}
%
with $\phi_\mu:=\nabla_\mu\phi$ and $\phi_{\mu\nu}:=\nabla_\mu\nabla_\nu\phi$.
In general, the Euler-Lagrange equations derived from this action, 
%
\begin{align}
  \mathcal{E}_A &:= \frac{1}{\sqrt{-g}}\sum_{j=0}(-1)^j\partial_{\mu_1}\cdots\partial_{\mu_j}\frac{\delta (\sqrt{-g}\mathcal{L}_{\rm DHOST})}{\delta\,\partial_{\mu_1}\cdots\partial_{\mu_j} A} = 0,
\label{eq:def_EA}
\end{align}
%
contain fourth-order time-derivatives of the scalar field.
To remove the ghost degree of freedom arising from the higher-derivative terms, we need to impose the degeneracy conditions \cite{Langlois:2015cwa,BenAchour:2016cay,Crisostomi:2016czh},
%
\begin{align}
&a_2=-a_1\,,\notag\\
&a_4=\frac{1}{8\left(f_2+a_2X\right)^2}\Bigl[
16Xa_2^2+4\left( 3f_2+16Xf_{2X}\right) a_2^2
+\left( 16X^2f_{2X}-12Xf_2\right) a_3a_2-X^2f_2a_3^2+48f_2f_{2X}^2
\notag \\ &
\quad+16f_{2X}\left( 3f_2+4Xf_{2X}\right) a_2+8f_2\left( Xf_{2X}-f_2\right)a_3
\Bigr]\,
,\label{eq:degeneracy_DHOST}\\
&a_5=\frac{\left( 4f_{2X}+2a_2+Xa_3\right)\left( -2a_2^2+3Xa_2a_3-4f_{2X}a_2+4f_2a_3\right)}{8\left(f_2+Xa_2\right)^2}
,
\end{align}
%
where the subscript $X$ denotes the derivatives with respect to $X$, e.g. $f_{2X} := df_2/dX$.

Assuming the flat Friedmann-Lema\^itre-Robertson-Walker metric, $ds^2 = -N^2dt^2 + a^2\delta_{ij}dx^idx^j$, 
we obtain the background equations,
%
\begin{align}
\mathcal{E}_N &= \rho, \quad -\frac{a}{3}\mathcal{E}_a = p
, \quad \mathcal{E}_\phi = 0. \label{eq:background_Na}
\end{align}
%
The right-hand sides of these equations are obtained by taking the GR limit in which $P=Q=a_i=0$ and $f_2=\Mpl^2/2$.
We omit to show their explicit forms in the general case. Those in the limited model that we will mention are shown in Sec.~\ref{subsec:cCK}.
As for the matter content, we assume that it consists of cold dark matter (CDM), baryons, photons, and massless neutrinos.

\subsection{Perturbations}

In our Boltzmann solver, we employ the EFT approach parameterised by arbitrary functions of time.
We write the metric in the ADM form,
%
\begin{align}
 ds^2 = -N^2dt^2+h_{ij}(dx^i+N^idt)(dx^j+N^jdt).
\end{align}
%
Expanding the action (\ref{eq:DHOST_action}) for the the scalar perturbation, $\delta\phi(t,\xx) := \phi(t,\xx)-\phi_0(t)$,
and the metric perturbations, $\delta g_{\mu\nu}(t,\xx) = g_{\mu\nu}(t,\xx)-g^{(0)}_{\mu\nu}(t)$, 
we obtain their quadratic action. The action can be recast in the unitary gauge as
\cite{Langlois:2017mxy}
%
\begin{align}
 \delta_2S &= \frac{1}{2}\int\!dtd^3x\,\sqrt{h}NM^2
 \left\{
  \delta K_{ij}\delta K^{ij}
 - \left(1+\frac{2}{3}\alpha_{\rm L}\right)\delta K^2
 +(1+\alpha_{\rm T})\left({}^{(3)}R\frac{\delta\sqrt{h}}{a^3}+\delta_2{}^{(3)}R\right)
\right.\notag \\ &\left.
 + H^2\alpha_{\rm K}\delta N^2
 + 4H\alpha_{\rm B}\delta K\delta N
 + (1+\alpha_{\rm H})R\delta N
 + 4\beta_1\delta K\dot{\delta N}
 + \beta_2\dot{\delta N}^2
 + \frac{\beta_3}{a^2}(\partial\delta N)^2
 \right\}, \label{eq:quad_action}
\end{align}
%
where $M(t)$ is the effective Planck mass, $K_{ij}$ the extrinsic curvature, ${}^{(3)}R$ the Ricci scalar
on the spatial hypersurface and 
$\delta_2$ operator extracts the second-order terms of the metric perturbations.
We introduce eight time-dependent parameters characterising the effective quadratic Lagrangian.
They are labelled as $\{\alpha_{\rm L},\alpha_{\rm T},\alpha_{\rm K},\alpha_{\rm B},\alpha_{\rm H},\beta_1,\beta_2,\beta_3\}$.
In addition, we introduce a parameter characterising
the time-variation of the effective Planck mass,
%
\begin{align}
 \alpha_{\rm M}:=\frac{1}{H}\frac{d}{dt}\ln M^2\,.
\end{align}
%
We define the metric perturbations as
%
\begin{align}
&N = 1+\delta N, \quad 
N^i=\delta^{ij}\partial_i\psi, \quad 
h_{ij}=a^2e^{2\zeta}\delta_{ij}+a^2\left(\partial_i\partial_i-\frac{1}{3}\delta_{ij}\partial^k\partial_k\right)E.
\label{eq:unitary gauge}
\end{align}
%
The infinitesimal time transformation, $t \to t+\pi(t,\xx)$, induces the gauge transformation
for the perturbative variables,
%
\begin{align}
&\delta N \;\to\; \Psi = \delta N - \dot{\pi}, \quad \zeta \;\to\; \Phi = \zeta-H\pi, 
\quad \psi \;\to\; \xi = \psi+\frac{1}{a^2}\pi, \quad E \;\to\; E.
\end{align}
%
Imposing the gauge condition, $\xi = E = 0$, we can move to the Newtonian gauge.
As a result of the time-coordinate transformation, the homogeneous scalar field 
in the unitary gauge acquires the spatial dependence, $\phi(t) \to \phi_0(t) + \delta\phi(t,\xx)$.
Thus
we can identify the scalar perturbation as \cite{Bellini:2014fua}
%
\begin{align}
  \pi := -\frac{\delta\phi}{\dot{\phi}_0}. \label{eq:def_pi}
\end{align}
%

The degeneracy conditions (\ref{eq:degeneracy_DHOST}) are recast as
%
\begin{align}
&\alpha_{\rm L}=0, \quad \beta_2=-6\beta_1^2, \quad \beta_3=-2\beta_1\left[2(1+\alpha_{\rm H})+\beta_1(1+\alpha_{\rm T})\right].
\label{eq:degeneracy_EFT}
\end{align}
%
They reduce the number of the independent EFT parameters to six, $\alpha_{\rm K}, \alpha_{\rm B}, \alpha_{\rm T}, \alpha_{\rm M}, \alpha_{\rm H}$ and $\beta_1$.
For details on the evolution equations for $\Psi, \Phi$ and $\pi$, see Ref.~\cite{Hiramatsu:2020fcd}. 
Although these equations still contain the higher-derivative terms up to the fourth-order,
they can be eliminated by a linear combination of the equations thanks to the degeneracy condition (\ref{eq:degeneracy_EFT}).
As the matter sector minimally couples to gravity, we obtain the same evolution equations for the perturbations of
CDM, baryons, photons and massless neutrinos as in GR.

\section{Model and setup}
\label{sec:model}

\subsection{EFT approach}

In the EFT approach, we can treat the background and perturbative quantities separately.
For instance, in Ref.~\cite{DAmico:2016ntq}, the authors assume the $\Lambda$CDM background and 
study the impact of $\alpha_{\rm H}$ on the angular power spectrum of the CMB anisotropies.
To study the significance of the parameter $\beta_1$ characterising the DHOST theories,
we also assume the $\Lambda$CDM background and consider a simple case with $\beta_1 \ne 0$, $\alpha_{\rm K} \ne 0$ and $\alpha_{\rm H}=\alpha_{\rm M}=\alpha_{\rm T}=\alpha_{\rm B}=0$.
Following the literature \cite{DAmico:2016ntq}, we assume that the EFT parameters behave as
%
\begin{align}
\beta_1(t) = \beta_{1,0}\frac{\Omega_{\rm DE}(t)}{\Omega_{\rm DE}(t_0)}, \quad
\alpha_{\rm K}(t) = \alpha_{{\rm K},0}\frac{\Omega_{\rm DE}(t)}{\Omega_{\rm DE}(t_0)},
\label{eq:scaling}
\end{align}
%
where $t_0$ is the present time and $\Omega_{\rm DE} := 1-\Omega_{\rm m}-\Omega_{\rm r}$ is the fractional energy density of the dark energy
with $\Omega_{\rm m}$ and $\Omega_{\rm r}$ being the fractional energy densities of the non-relativistic and relativistic matter contents, respectively.
We fix $\alpha_{{\rm K},0}=1$, and focus only on  the parameter $\beta_{1,0}$.
In this setup, we have to impose $\beta_{1,0}>0$ to avoid the ghost instability and the gradient instability of the curvature perturbations.


\subsection{Constrained Crisostomi-Koyama (cCK) model}
\label{subsec:cCK}

In order to treat the background and perturbative quantities consistently, as opposed to the EFT approach, 
we go back to the action (\ref{eq:DHOST_action}) with fixing the arbitrary functions, $P, Q, f_2, a_1$ and $a_3$.
Note that $a_2, a_4$ and $a_5$ are determined from Eq.~(\ref{eq:degeneracy_DHOST}).
One simple parameterisation has been proposed by Crisostomi and Koyama (CK) in Ref.~\cite{Crisostomi:2017pjs}, given by
%
\begin{align}
 P &= c_2X, \quad
 Q = \frac{c_3}{\Lambda^3}X, \quad
 f_2 = \frac{\Mpl^2}{2}+ c_4\frac{X^2}{\Lambda^6}, \quad
 a_1=0, \quad a_3 = -\frac{\beta+8c_4}{\Lambda^6},
\label{eq:action_CK}
\end{align}
%
where $c_2, c_3, c_4$ and $\beta$ are constant parameters.
This setup ensures that the speed of gravitational waves is the same as that of light, resulting from
the recent observation of the gravitational waves from a neutron star merger by the LIGO/VIRGO collaborations (GW170817) \cite{LIGOScientific:2017vwq,LIGOScientific:2017zic,LIGOScientific:2017ync}.

To prevent the gravitational waves from decaying into dark energy during their propagation, we have to require 
$a_3=0$ \cite{Creminelli:2018xsv}.
This requirement, as well as the exact luminality of the gravitational waves, results from 
GW170817 at $z\lesssim 0.01$, so it is not necessarily imposed on physics at the CMB scale, $z\sim 1100$ \cite{deRham:2018red}.
Nevertheless, in this paper, we impose the condition, $a_3=0$, to reduce the number of free parameters. As a result,
$\beta$ is determined from $c_4$, $\beta = -8c_4$. Moreover, we set $c_2=1$ to fix the normalisation of
the scalar field. Eventually, the free parameters are only $c_3$ and $c_4$.
We dub this subclass of the CK model as the constrained CK (cCK) model.
In this setup, $c_3>0$ and $c_4>0$ are required to guarantee the existence of an attractor solution for the background
scalar field \cite{Crisostomi:2018bsp}.

As the shift-symmetry holds in the action (\ref{eq:def_action}) with Eq.~(\ref{eq:action_CK}), 
the background equations (\ref{eq:background_Na}) 
do not depend on the scalar field itself, $\phi_0$.
Hence we can write them in terms of its time-derivative, $\chi:=\dot{\phi}_0$:
%
\begin{align}
&U_1(\chi,a)\dot{\chi} +U_2(\chi,a) = 0, \label{eq:CK_background_evol1}\\
&\frac{\dot{a}}{a} = V_1(\chi)\dot{\chi} + V_2(\chi,a),\label{eq:CK_background_evol2}
\end{align}
%
where $V_i$ and $U_i$ are shown in Appendix \ref{appsec:back_cCK}.
Note that, in the limit where $c_3,c_4 \to 0$, the model becomes 
equivalent to the k-essence model consisting of only $P$ term \cite{Armendariz-Picon:2000nqq,Armendariz-Picon:2000ulo}, not the $\Lambda$CDM model. 
According to Ref.~\cite{Hiramatsu:2020fcd}, we can easily read the EFT parameters as functions of $\chi$ and $a$.
We show them in Appendix \ref{appsec:alpha_param}.

\subsection{Setup for MCMC simulations}
\label{subsec:setup}

We perform the MCMC simulations in the EFT approach
and the cCK model as well as in the $\Lambda$CDM model as a reference. We use the Planck 2018 TTTEEE$+$lowE likelihoods provided by the Planck collaboration~\cite{Planck_Legacy_Archive}\footnote{The likelihood files provided by Planck are {\tt commander\_dx12\_v3\_2\_29.clik} for $TT$ in $2\leq \ell\leq 29$; {\tt plik\_rd12\_HM\_v22b\_TTTEEE.clik} for $TT+TE+EE$ in $30\leq \ell\leq 2508$; and {\tt simall\_100x143\_offlike5\_EE\_Aplanck\_B.clik} for $EE$ in $2\leq \ell \leq 29$.}.
To compute the angular power spectra, $C^{TT}_\ell, C^{TE}_\ell$ and $C^{EE}_\ell$, including the weak lensing effects, we use the Boltzmann solver developed in Ref.~\cite{Hiramatsu:2020fcd}.
The nuisance parameters are generated according to Ref.~\cite{Planck:2019nip}.

In the $\Lambda$CDM model, we choose the cosmological parameters $\{B_{\rm s}, n_{\rm s}, h, \omega_{\rm c}, \omega_{\rm b}, \tau\}$ to be 
varied in the MCMC simulations, where $B_{\rm s} := e^{-2\tau}A_{\rm s}$ is the rescaled amplitude of the primordial curvature perturbations, 
$n_s$ the spectral index, 
$\omega_i := h^2\Omega_i$ the fractional energy density of the matter for $i={\rm c}, {\rm b}, \gamma$ and $\nu$, 
$h$ the reduced Hubble parameter, and $\tau$ the optical depth.
As the amplitude of the curvature perturbation is highly degenerate with the optical depth,
we use the Gaussian prior with $\tau = 0.054\pm 0.023$. 
In addition to the six parameters, we vary $\beta_{1,0}$ in the EFT approach and $\{c_3,c_4\}$ in the cCK model.

\section{Results}
\label{sec:results}

\subsection{Best-fit parameters}

\begin{table}[!t]
\begin{tabular}{c|l|cl|cl}
\hline
\hline
                      & \multicolumn{1}{c}{$\Lambda$CDM}  & \multicolumn{2}{|c}{EFT}                          & \multicolumn{2}{|c}{cCK} \\
\hline
                      &~~~                                &~~             &~~~                               &~~$c_3$ &~~~$1.59_{-0.28}^{+0.26}$          \\[0.4em]     
                      &~~~                                &~~$\beta_{1,0}$&~~~$0.032_{-0.016}^{+0.013}$      &~~$c_4$ &~~~$<0.0088$                       \\[0.4em] \hline
$10^{9}B_s$           &~~~$1.8868_{-0.0056}^{+0.0066}$    &               &~~~$1.8986_{-0.0048}^{+0.010}$    &        &~~~$1.8816_{-0.0050}^{+0.0055}$    \\[0.4em]
$n_s$                 &~~~$0.9697_{-0.0027}^{+0.0042}$    &               &~~~$0.9735_{-0.0051}^{+0.0025}$   &        &~~~$0.9909_{-0.0030}^{+0.0028}$    \\[0.4em]
$h$                   &~~~$0.6782_{-0.0056}^{+0.0038}$    &               &~~~$0.6760_{-0.0073}^{+0.0048}$   &        &~~~$0.8043_{-0.0058}^{+0.0040}$    \\[0.4em]
$\omega_{\rm c}$      &~~~$0.11878_{-0.00064}^{+0.0015}$  &               &~~~$0.1194_{-0.0011}^{+0.0016}$   &        &~~~$0.11523_{-0.00072}^{+0.0011}$  \\[0.4em]
$\omega_{\rm b}$      &~~~$0.02177_{-0.00013}^{+0.00014}$ &               &~~~$0.02175_{-0.00017}^{+0.00012}$&        &~~~$0.02218_{-0.00012}^{+0.00013}$ \\[0.4em]
$\tau$                &~~~$0.0489_{-0.0045}^{+0.0058}$    &               &~~~$0.0494_{-0.0045}^{+0.0061}$   &        &~~~$0.0423_{-0.0071}^{+0.0044}$    \\[0.4em]
\hline
$10^{9}A_s$           &~~~$2.081_{-0.018}^{+0.025}$       &               &~~~$2.106_{-0.026}^{+0.018}$      &        &~~~$2.045_{-0.026}^{+0.021}$       \\[0.4em]
$\Omega_{\rm c}$      &~~~$0.2569_{-0.0028}^{+0.0088}$    &               &~~~$0.2616_{-0.0062}^{+0.0091}$   &        &~~~$0.1782_{-0.0029}^{+0.0041}$    \\[0.4em]
$\Omega_{\rm b}$      &~~~$0.04726_{-0.00031}^{+0.00074}$ &               &~~~$0.04757_{-0.00052}^{+0.00079}$&        &~~~$0.03432_{-0.00030}^{+0.00039}$ \\[0.4em]
\hline
$\ln\mathcal{L}$      &~~~$-1432$                         &               &~~~$-1428$                        &        &~~~$-1463$                         \\[0.4em]
\hline
\hline
\end{tabular}
\caption{The best-fit parameters and their 68$\%$ confidence intervals in the $\Lambda$CDM model, 
EFT and cCK model. The three parameters in the lower part, $A_s, \Omega_{\rm c}$ and $\Omega_{\rm b}$, 
are derived from the parameters above. The inequality indicates that we obtain only the upper limit. }
\label{tab:bestfit}
\end{table}

The best-fit parameters and their 68$\%$ confidence intervals are presented in Table~\ref{tab:bestfit}.
The first two rows are the extra parameters in the EFT approach and the cCK model, and $A_{\rm s}, \Omega_{\rm c}$ and $\Omega_{\rm b}$ are derived from the above parameters.
In the $\Lambda$CDM model, we almost reproduce the Planck 2018 results \cite{Planck:2018vyg}.
In the bottom line, we show the values of the likelihood for the best-fit parameters. In both the EFT approach and the cCK model, 
the fitting is not so improved, although we introduce extra parameters in addition to the standard six parameters.

\begin{figure}[!t]
\begin{tikzpicture}
\node (img) {\includegraphics[width=8cm]{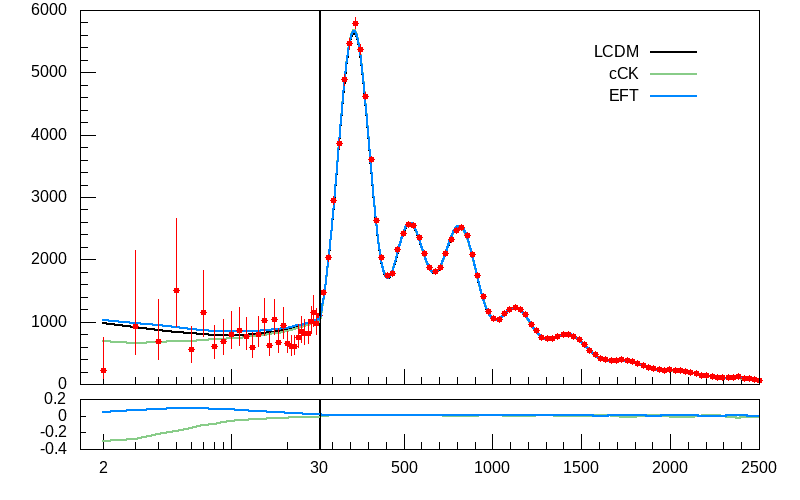}};
\node[below=of img, node distance=0cm, yshift=1.2cm, xshift=0.4cm, font=\small] {$\ell$};
\node[left=of img, node distance=0cm, rotate=90, anchor=center, yshift=-1.1cm, xshift=0.3cm, font=\small] {$D_{\ell}^{\rm TT}~[(\mu K)^2]$};
\node[left=of img, node distance=0cm, rotate=90, anchor=center, yshift=-1.1cm, xshift=-1.7cm, font=\small] {$\Delta^{\rm TT}$};
\end{tikzpicture}
\begin{tikzpicture}
\node (img) {\includegraphics[width=8cm]{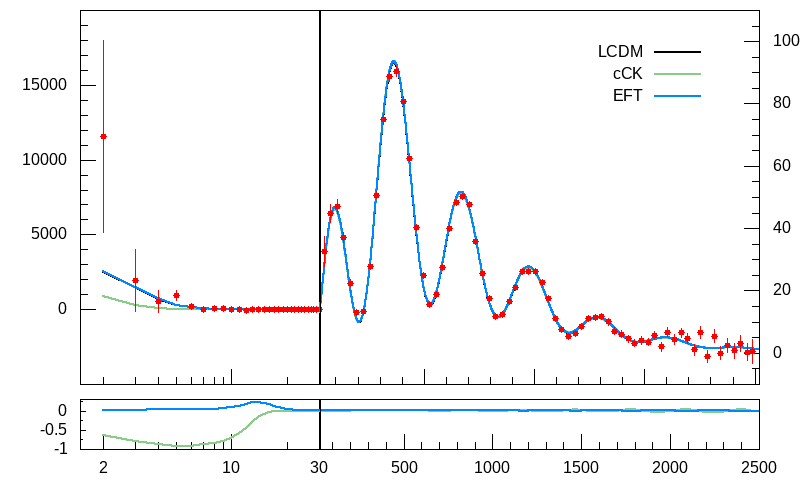}};
\node[below=of img, node distance=0cm, yshift=1.2cm, xshift=0.4cm, font=\small] {$\ell$};
\node[left=of img, node distance=0cm, rotate=90, anchor=center, yshift=-1.1cm, xshift=0.3cm, font=\small] {$D_{\ell}^{\rm EE}~[(\mu K)^2]$};
\node[left=of img, node distance=0cm, rotate=90, anchor=center, yshift=-1.1cm, xshift=-1.7cm, font=\small] {$\Delta^{\rm EE}$};
\end{tikzpicture}
\caption{Angular power spectra, $D^X_{\ell} := \ell(\ell+1)C^X_{\ell}/(2\pi)$,  of the temperature fluctuations ({\it left}) and the E-mode polarisation ({\it right}) with the best-fit parameters in the $\Lambda$CDM model ({\it black}), EFT ({\it cyan}), and cCK model ({\it green}). The elongated panels below the spectra show the fractional deviation from the $\Lambda$CDM case, $\Delta^{X} := (C_\ell^{X}-C_\ell^{X(\Lambda{\rm CDM})})/C_\ell^{X(\Lambda{\rm CDM})}$ for $X={\rm TT}, {\rm EE}$. Notice that, in the right panel, the vertical scale for $\ell\geq 30$ shown in the right vertical axis is different from that for $\ell < 30$ shown in the left vertical axis. 
}
\label{fig:bestfit}
\end{figure}
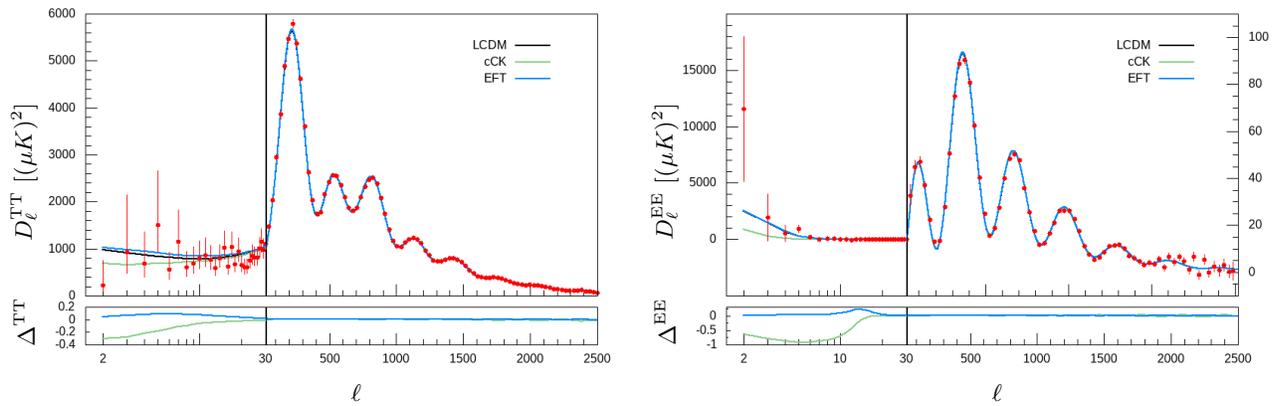

The rescaled angular power spectra, $D^X_{\ell} := \ell(\ell+1)C^X_{\ell}/(2\pi)$, for $X={\rm TT}, {\rm EE}$, are shown in Fig.~\ref{fig:bestfit}. 
The red points with error bars are the binned data from Planck 2018 results \cite{Planck:2018vyg}. The black, cyan and green lines are the best-fit curves in the $\Lambda$CDM model, EFT and the cCK model, respectively. The extra parameters, $\beta_{1,0}$ and $\{c_3,c_4\}$, affect the angular power spectra only on large scales, $\ell \lesssim 30$, as clarified in our previous paper \cite{Hiramatsu:2020fcd}.

As a result of the MCMC simulations, we obtain
%
\begin{align}
\beta_{1,0} = 0.032_{-0.016}^{+0.013} \quad ({\rm EFT}); \quad
c_3=1.59_{-0.28}^{+0.26}, \quad (0<)~c_4 < 0.0088 \quad ({\rm cCK}).
\end{align}
%
These are the first results ever to put constraints on the model parameters in the DHOST theories from CMB observations. 

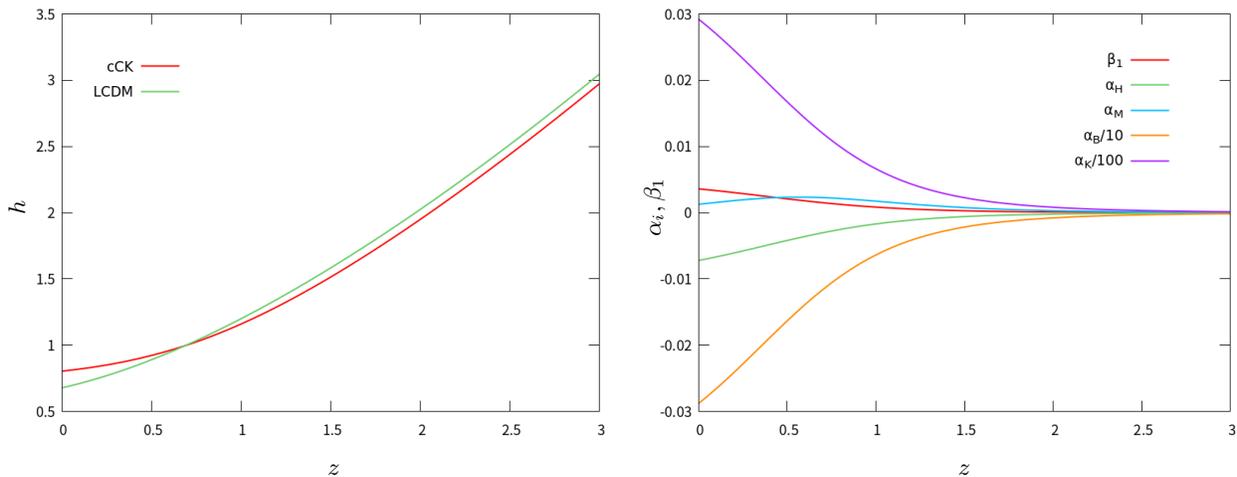
\begin{figure}[!t]
\begin{tikzpicture}
\node (img) {\includegraphics[width=8cm]{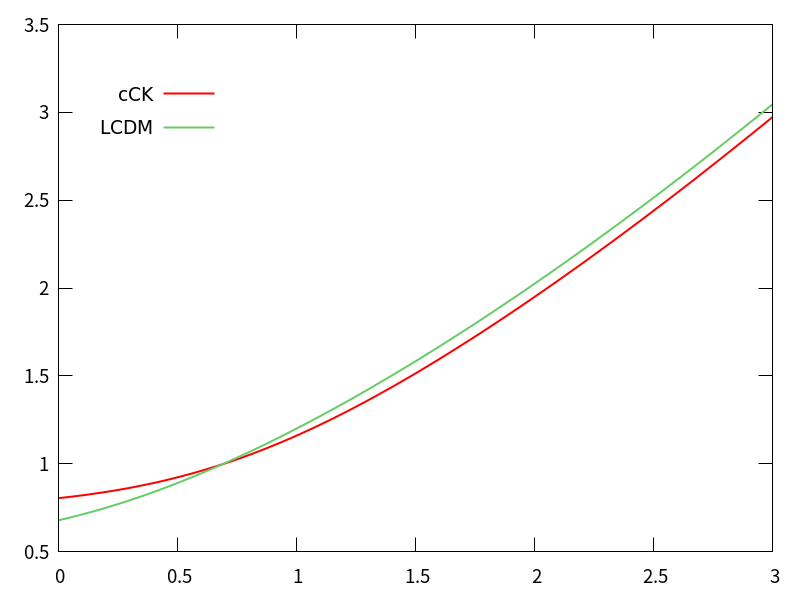}};
\node[below=of img, node distance=0cm, yshift=1.0cm, xshift=0.2cm, font=\small] {$z$};
\node[left=of img, node distance=0cm, rotate=90, anchor=center, yshift=-1.1cm, xshift=0.2cm, font=\small] {$h$};
\end{tikzpicture}
\begin{tikzpicture}
\node (img) {\includegraphics[width=8cm]{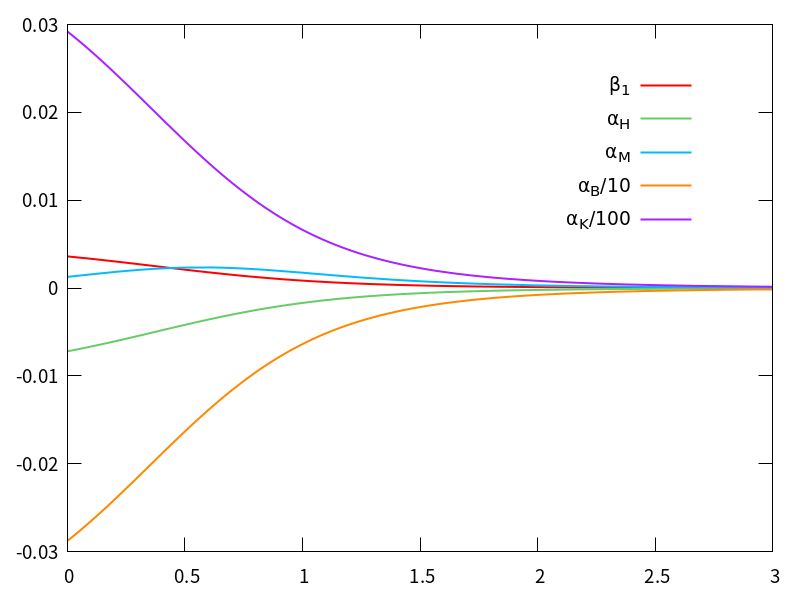}};
\node[below=of img, node distance=0cm, yshift=1.0cm, xshift=0.2cm, font=\small] {$z$};
\node[left=of img, node distance=0cm, rotate=90, anchor=center, yshift=-1.2cm, xshift=0.2cm, font=\small] {$\alpha_i, \beta_1$};
\end{tikzpicture}
\caption{The time-evolution of the reduced Hubble parameter in the cCK model and the $\Lambda$CDM model ({\it left}), and the EFT parameters as functions of the redshift with the best-fit values in the cCK model ({\it right}). As the parameters $\alpha_{\rm B}$ and $\alpha_{\rm K}$ are relatively larger than the others, we rescale them as $\alpha_{\rm B}/10$ and $\alpha_{\rm K}/100$.}
\label{fig:background_alpha_cCK}
\end{figure}

In the cCK model, we find that the reduced Hubble parameter at the present time, $h=0.8043$, is a little bit larger than the best-fit value in the $\Lambda$CDM model, $h=0.6782$. To see why it is enhanced, we show the time-evolution of the reduced Hubble parameter
in the left panel of Fig.~\ref{fig:background_alpha_cCK}. 
The Hubble parameter is basically smaller than that in the $\Lambda$CDM model at the high redshift.
Its decreasing rate, however, suddenly slows down at $z\sim 1$ and the present value becomes larger than that in the $\Lambda$CDM model. As a result, the comoving distance to the last-scattering surface does not change in both models. 
That is why the present value of the Hubble parameter in the cCK model is larger than that in the $\Lambda$CDM model.

In the right panel of Fig.~\ref{fig:background_alpha_cCK}, we show the time-evolution of the EFT parameters in the cCK model with the best-fit parameters.
In contrast to the EFT approach that we consider here, all the EFT parameters, except for $\alpha_{\rm T}$ being zero, have non-zero values. In particular, $\alpha_{\rm M}$ changes non-monotonically in time,
as pointed out in our previous paper \cite{Hiramatsu:2020fcd}.
The physical effects of the EFT parameters on the angular power spectra are more or less degenerate with each other. 
Therefore these parameters can take large values since their influence on the angular power spectra can be cancelled.
However, in the cCK model, $\beta_1$ and $\alpha_{\rm H}$ always satisfy the relation, 
$\beta_1=-\alpha_{\rm H}/2$ [see Eq.~(\ref{eq:cCK_beta_H})], and thus the cancellation does not work. 
Hence we can put a relatively strong constraint on the parameter $c_4$, which controls the significance of $\beta_1$.
We will see the impact of this parameter on the angular power spectra in Sec.~\ref{subsec:c3c4}.

The parameter $\beta_{1,0}$ has been constrained by the other astrophysical observations.
The long-term observation of the Hulse-Taylor pulsar gives a bound on the effective gravitational coupling constant for the gravitational waves.
That puts a constraint on $\beta_{1,0}$ as $|\beta_{1,0}|\lesssim \mathcal{O}(10^{-3})$ \cite{Hirano:2019scf}. The Solar System test provides a tighter constraint on $\beta_{1,0}$ as $0\leq \beta_{1,0}\lesssim 10^{-5}$ \cite{Crisostomi:2019yfo}. These are stronger constraints than our present result based only on the cosmological-scale observation. 

\subsection{Correlation between parameters}

\begin{figure}[!t]
\begin{tikzpicture}
\node (img) {\includegraphics[width=6.6cm]{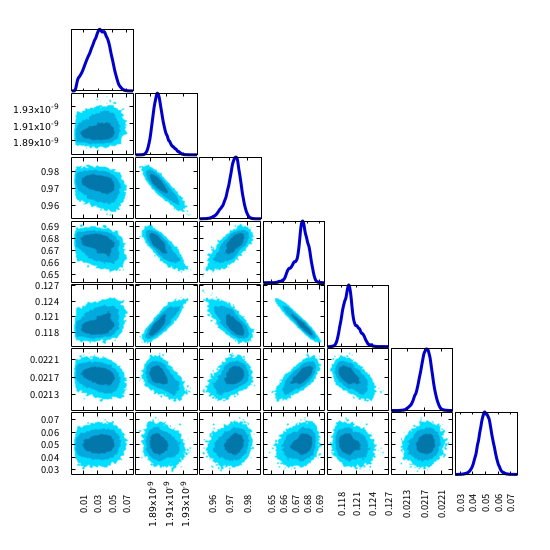}};
\node[below=of img, node distance=0cm, yshift=1.2cm, xshift=-2.4cm, anchor=west, font=\tiny] {$\beta_{1,0}\hspace{4.0mm}B_s\hspace{4.8mm}n_s\hspace{4.8mm}h\hspace{6.0mm}\omega_{\rm c}\hspace{5.5mm}\omega_{\rm b}\hspace{4.0mm}\tau$};
\node[left=of img, node distance=0cm, rotate=90, anchor=west, yshift=-1.0cm, xshift=-2.2cm, font=\tiny] {$\tau\hspace{6.0mm}\omega_{\rm b}\hspace{4.5mm}\omega_{\rm c}\hspace{5.0mm}h\hspace{6.0mm}n_s\hspace{4.5mm}B_s\hspace{4.0mm}\beta_{1,0}$};
\end{tikzpicture}
\begin{tikzpicture}
\node (img) {\includegraphics[width=7.2cm]{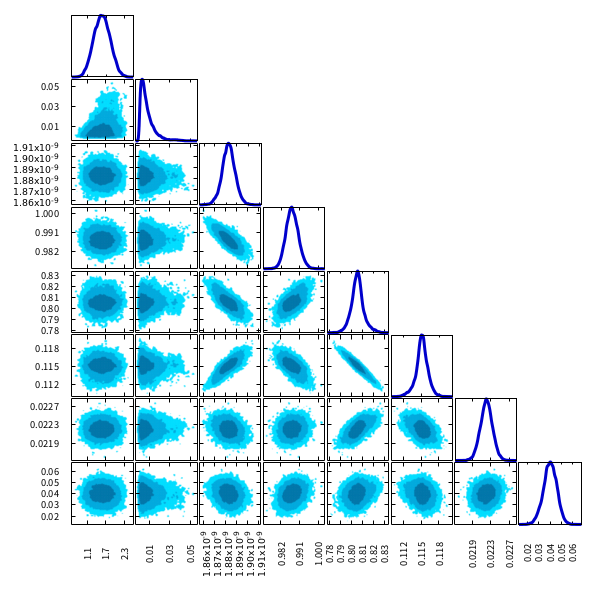}};
\node[below=of img, node distance=0cm, yshift=1.2cm, xshift=-2.6cm, anchor=west, font=\tiny] {$c_3\hspace{5mm}c_4\hspace{5.5mm}B_s\hspace{4.5mm}n_s\hspace{5.5mm}h\hspace{6.0mm}\omega_{\rm c}\hspace{4.5mm}\omega_{\rm b}\hspace{4.5mm}\tau$};
\node[left=of img, node distance=0cm, rotate=90, anchor=west, yshift=-1.0cm, xshift=-2.5cm, font=\tiny] {$\tau\hspace{6.0mm}\omega_{\rm b}\hspace{4.5mm}\omega_{\rm c}\hspace{5.0mm}h\hspace{5.5mm}n_s\hspace{5.0mm}B_s\hspace{5.0mm}c_4\hspace{5mm}c_3$};
\end{tikzpicture}
\begin{tikzpicture}
\node (img) {\includegraphics[width=6.0cm]{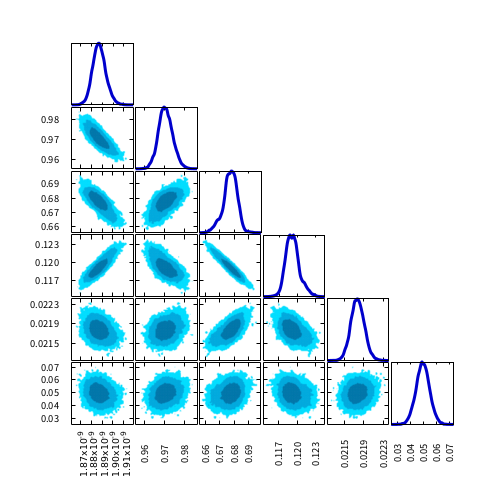}};
\node[below=of img, node distance=0cm, yshift=1.2cm, xshift=-2.0cm, anchor=west, font=\tiny] {$B_s\hspace{4.5mm}n_s\hspace{5mm}h\hspace{6.0mm}\omega_{\rm c}\hspace{5.0mm}\omega_{\rm b}\hspace{5.0mm}\tau$};
\node[left=of img, node distance=0cm, rotate=90, anchor=west, yshift=-1.2cm, xshift=-1.9cm, font=\tiny] {$\tau\hspace{6.0mm}\omega_{\rm b}\hspace{4.5mm}\omega_{\rm c}\hspace{5.0mm}h\hspace{5.5mm}n_s\hspace{5.0mm}B_s$};
\end{tikzpicture}
\caption{Confidence contours at 68\%, 95\% and 99\% confidence levels in the EFT approach ({\it top-left}), the cCK model ({\it top-right}) and the $\Lambda$CDM model ({\it bottom}).
We also show the marginalised 1D distributions for each parameter in the diagonal side.}
\label{fig:contour}
\end{figure}
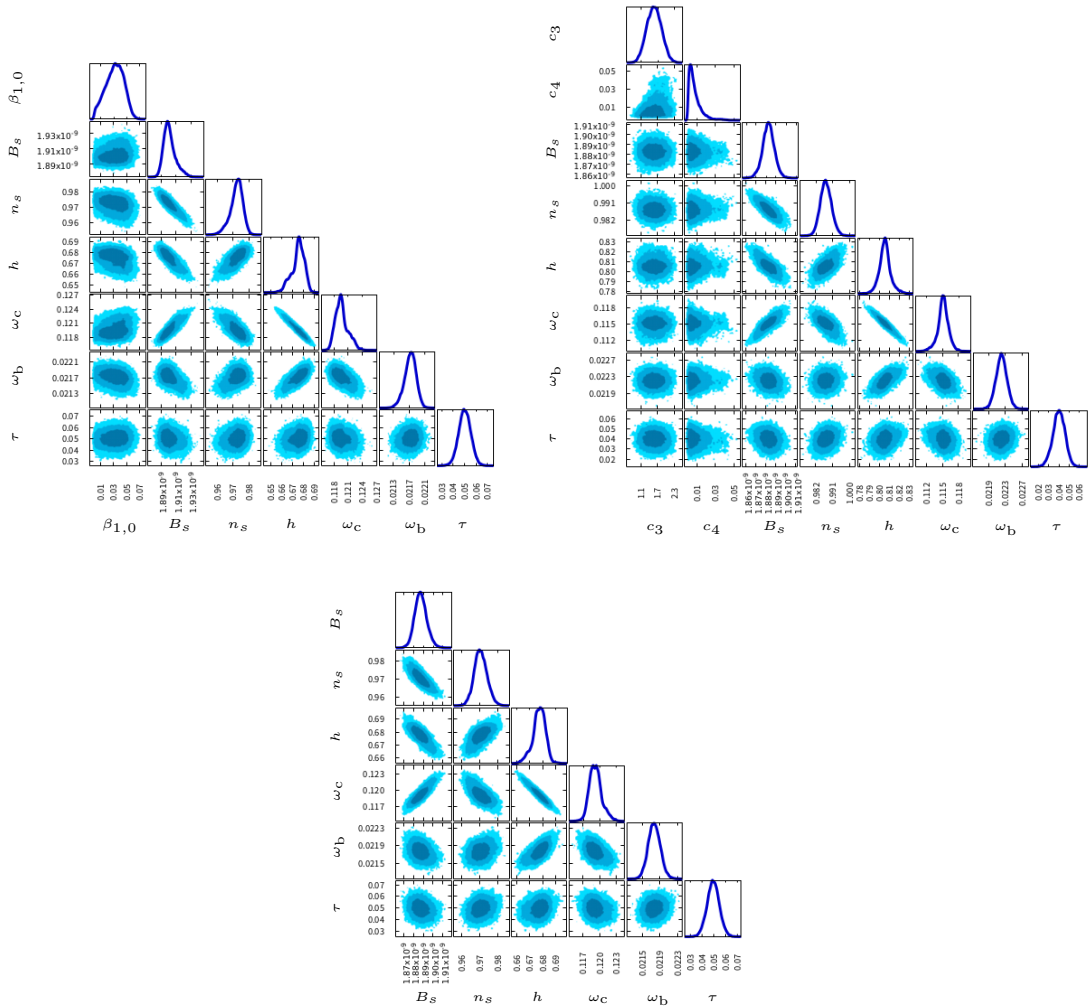

In Fig.~\ref{fig:contour}, we present the 2D contour plots of the multivariate 
probability distributions with the kernel density estimation technique \cite{Sheather:2004}
in EFT ({\it top-left}), cCK ({\it top-right}) and $\Lambda$CDM ({\it bottom}) models.
The colour contours indicate the confidence regions at 68\%, 95\% and 99\% confidence levels, and we also show the marginalised 1D distributions for each parameter in the diagonal side.
In the EFT approach and the cCK model, the extra parameters, $\beta_{1,0}$ (EFT) and $\{c_3,c_4\}$ (cCK), are not strongly correlated with the other six parameters.
They modify the gravity potentials at the late time, leading to an extra contribution to the integrated Sachs-Wolfe effect on the angular power spectrum of the temperature anisotropies. This effect is almost independent to the effects controlled by the other six parameters.

\subsection{Impact of $c_3$ and $c_4$ parameters in the cCK model}
\label{subsec:c3c4}

\begin{figure}[!t]
\begin{tikzpicture}
\node (img) {\includegraphics[width=8cm]{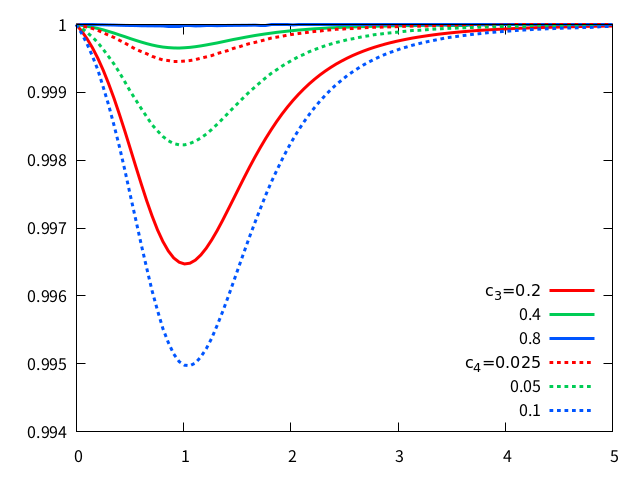}};
\node[below=of img, node distance=0cm, yshift=1cm, xshift=0.25cm] {$z$};
\node[left=of img, node distance=0cm, rotate=90, anchor=center, yshift=-0.7cm] {$H(z)/H_{\rm cCK,best}(z)$};
\end{tikzpicture}
\begin{tikzpicture}
\node (img) {\includegraphics[width=8cm]{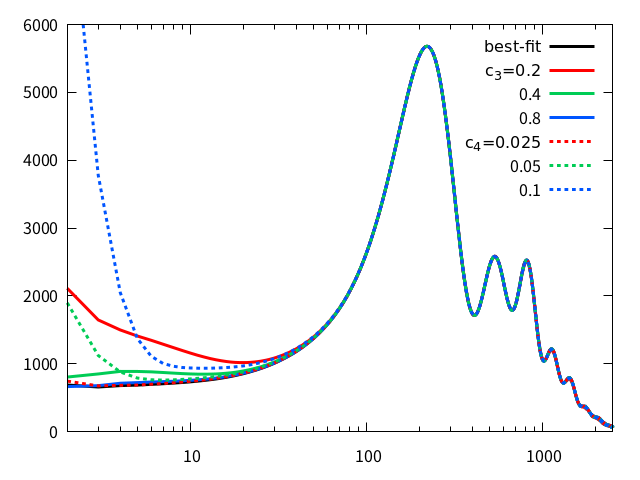}};
\node[below=of img, node distance=0cm, yshift=1.2cm] {$\ell$};
\node[left=of img, node distance=0cm, rotate=90, anchor=center, yshift=-0.7cm, xshift=0.2cm] {$\ell(\ell+1)C_{\ell}^{TT}/(2\pi)~[(\mu K)^2]$};
\end{tikzpicture}
\caption{The left panel shows the Hubble parameter with various $c_3$ ({\it solid}) and various $c_4$ ({\it dotted}) in the cCK model, normalised by the best-fit case. The other parameters are fixed to be the best-fit ones. Note that the best-fit values of $c_3$ and $c_4$ are $(c_3,c_4)=(1.59,0)$. The right panel shows the angular power spectrum with the same variations of $c_3$ ({\it solid}) and $c_4$ ({\it dotted}). }
\label{fig:Cl_cCK}
\end{figure}
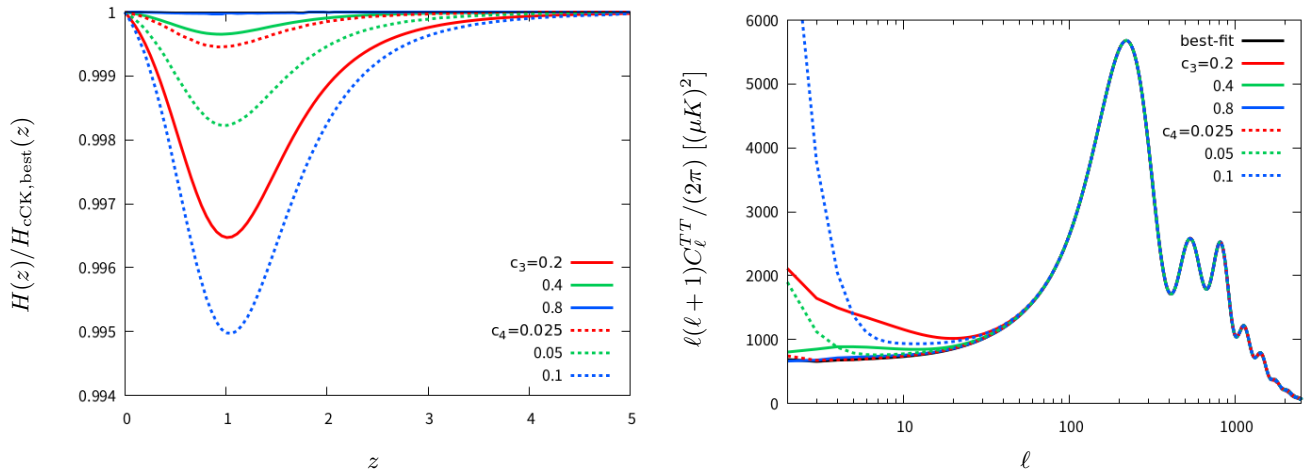

In the left panel of Fig.~\ref{fig:Cl_cCK}, we show the Hubble parameter with various $c_3$ ({\it solid}) and $c_4$ ({\it dotted}). The vertical axis is the Hubble parameter normalised by that with the best-fit case, $(c_3,c_4) = (1.59, 0)$. Although we vary these parameters in wider ranges than their confidence intervals, the Hubble parameter changes
only by at most $0.5\%$. This test implies that $c_3$ and $c_4$ are not well constrained from 
the time-evolution of the Hubble parameter. 
On the other hand, the same variations of these parameters significantly alter the angular power spectrum $C^{\rm TT}_\ell$ as shown in the right panel of Fig.~\ref{fig:Cl_cCK}. 
From this fact, both $c_3$ and $c_4$ cannot deviate from the best-fit values; otherwise, the angular power spectrum on the large scales are too enhanced and thus such parameter regions are obviously ruled out.

\section{Conclusion}
\label{sec:conclusion}

We investigate the impact of the model parameters characterising the Type-I quadratic DHOST theories on the CMB angular power spectra, and estimate their viable parameter regions using the MCMC method with TTTEEE$+$low-E likelihood based on the Planck 2018 observation. We employ two approaches, the EFT approach with a fixed background, and the DHOST theory with the parameterisation proposed by Crisostomi and Koyama in which we also solve the background equations for the scalar field and the scale factor.

First, we consider the EFT action that is characterised by six time-dependent parameters, $\alpha_{\rm K}, \alpha_{\rm B}, \alpha_{\rm T}, \alpha_{\rm M}, \alpha_{\rm H}$ and $\beta_1$.
In particular, the parameter $\beta_1$ characterises the Type-I DHOST theories beyond the GLPV theory.
For simplicity, we focus on a limited case where $\alpha_{\rm B}=\alpha_{\rm T}=\alpha_{\rm M}=\alpha_{\rm H} = 0$.
The remaining parameters $\alpha_{\rm K}(t)$ and $\beta_1(t)$ are assumed to be proportional to the fractional amount of the dark energy, i.e. $\alpha_{\rm K} = \alpha_{\rm K,0}\Omega_{\rm DE}(t)/\Omega_{\rm DE}(t_0)$ and $\beta_1 = \beta_{1,0}\Omega_{\rm DE}(t)/\Omega_{\rm DE}(t_0)$ where $t_0$ is the present time.
We fix $\alpha_{K,0}=1$ and focus on the unique constant parameter $\beta_{1,0}$ in the $\Lambda$CDM background.
As a result of the MCMC analysis varying the parameter $\beta_{1,0}$ as well as the six standard cosmological parameters, we obtain $\beta_{1,0}=0.032_{-0.016}^{+0.013}$ at the 68\% confidence level.

Next, to treat both the background and perturbations consistently, we test a DHOST theory 
with specifying the arbitrary functions given in Eq.~(\ref{eq:action_CK}).
We solve the scalar field equation and the modified Friedmann equation written in terms of the arbitrary functions to determine the background scalar field and the cosmic expansion history.
This treatment is totally different from the EFT approach in which we can assume an arbitrary cosmic expansion 
history, such as the $\Lambda$CDM background, bearing no relation to the EFT parameters.
We define a subclass theory in this model by setting $c_2=1$ and $\beta=-8c_4$ in Eq.~(\ref{eq:action_CK}) to reduce
the number of the model parameters. We dub this model as cCK model in the main text.
Note that this model ensures the luminality of the gravitational waves and their stability during propagation. As a result of the MCMC analysis varying the standard six 
cosmological parameters 
and the extra parameters $c_3$ and $c_4$, we obtain $c_3 = 1.59^{+0.26}_{-0.28}$ and $(0<)~c_4<0.0088$ at the 68\% confidence level. Note that we assume flat priors on $c_3>0$ and $c_4>0$ to ensure the existence of 
the attractor solution for the background quantities. 

We have put constraints on the model parameters in the subclass of DHOST theories using only the Planck 2018 likelihoods. They would be highly improved if we take the amplitude of the matter fluctuation, $\sigma_8$, into account combining the observational results from the large-scale structure survey. 
We leave it for the future work.

\acknowledgments{
 We would like to thank Katsuki Aoki for fruitful discussion during his stay in Rikkyo University.
 We thank Atsushi Taruya and Daisuke Yamauchi for giving helpful comments on our main results at 77th annual meeting of the Japan Physics Society,
 and also thank Kiyotomo Ichiki, Takahiro Nichimichi, Tomo Takahashi, and Tsutomu Kobayashi for useful discussion.
 This work was supported in part by JSPS KAKENHI Grants No. JP21K03559.
}

\appendix

\section{Background equations in cCK model}
\label{appsec:back_cCK}

The background equations in the cCK model are given in Eqs.~(\ref{eq:CK_background_evol1})(\ref{eq:CK_background_evol2}), whose 
coefficients, $U_i$ and $V_i$, are defined as
\begin{align}
U_1(\chi,a) &= 
2-2 c_{3} S \chi-8\left(3 c_{3}^2-7 c_{4}\right) \chi^4-12 c_{3} c_{4} S \chi^5+24c_{4}^2 (1-3 w) \rho \chi^6-4 c_{4} \left(15 c_{3}^2-34 c_{4}\right) \chi^8
\notag \\ &
-32 c_{3} c_{4}^2 S\chi^9+80 c_{4}^3 (1-3 w) \rho \chi^{10}-192 c_{3}^2 c_{4}^2 \chi^{12}-16 c_{3} c_{4}^3 S \chi^{13}+32 c_{4}^4 (1-3 w) \rho \chi^{14}
\notag \\ &
-48 c_{4}^3 \left(3 c_{3}^2-10 c_{4}\right) \chi^{16}
,\\
U_2(\chi,a) &= 
\chi(1+2c_{4}\chi^4)\left(S+6 c_{3} \chi^3-\left(3 c_{3}^2+2 c_{4}\right) S \chi^4+6 c_{3} c_{4} (1-w) \rho \chi^5
-6 \left(3 c_{3}^3+2 c_{3} c_{4}\right) \chi^7
\right. \notag \\ & \left.
-6 c_{4}\left(c_{3}^2-2 c_{4}\right) S \chi^8+12 c_{3} c_{4}^2 (1-5 w) \rho \chi^9-12 c_{3}c_{4} \left(3 c_{3}^2-10 c_{4}\right) \chi^{11}
\right)
,\\
V_1(\chi)   &=-\frac{4 c_{4} \chi^3}{1+2 c_{4} \chi^4}, \\ 
V_2(\chi,a) &=\frac{S+6 c_{3} \chi^3}{6-12 c_{4} \chi^4},
\end{align}
%
where $w:=(\rho_\gamma+\rho_\nu)/(3\rho)$ and 
%
\begin{align}
S &= 
2 \sqrt{3 (1-12 w) \rho +33 \chi ^2-6 c_{4} (1-6 w) \rho  \chi ^4+3 \left(3 c_{3}^2-10 c_{4}\right) \chi ^6+\frac{36\left(w \rho -\chi ^2\right)}{1+2 c_{4} \chi ^4}}.
\end{align}
%

\section{EFT parameters in the cCK model}
\label{appsec:alpha_param}

We can easily find the expressions of the EFT parameters in terms of the arbitrary functions of the DHOST theories \cite{Hiramatsu:2020fcd}.
In the cCK model, they are given as
%
\begin{align}
\beta_1 &= -\frac{1}{2}\alpha_{\rm H} = \frac{4 c_{4} \chi^4}{1+2 c_{4} \chi^4}
\label{eq:cCK_beta_H}
,\\
\alpha_{\rm M} &=\frac{8c_{4}\chi^3\dot{\chi} }{H(1+2 c_{4}\chi^4)}
,\\
\alpha_{\rm T} &= 0
,\\
\alpha_{\rm B} &=
\frac{\chi^3}{H \left(1+2 c_{4} \chi^4\right)^2} \left\{-(c_{3}+4 c_{4} H \chi) \left(1+2 c_{4} \chi^4\right)+4 c_{4} \left(5-2 c_{4} \chi^4\right) \dot{\chi}\right\}
,\\
\alpha_{\rm K} &= 
\frac{2 \chi^2}{H^2 \left(1+2 c_{4} \chi^4\right)^4} \left[-1+2 \chi \left\{6 c_{4} (6H^2 +5\dot{H}) \chi \left(1+2 c_{4} \chi^4\right)^3
+3 H\left(c_{3} \left(1+2 c_{4} \chi^4\right)^3+192 c_{4}^2 \chi^4 \left(1+c_{4} \chi^4\right) \left(1+2 c_{4} \chi^4\right) \dot{\chi}\right)
\right.\right. \notag \\ & \left.\left.
+c_{4} \chi^3 \left(-3+2 c_{4} \left(300 \dot{\chi}^2+\chi^4 \left(-3-2 c_{4} \chi^4+144c_{4} \left(2+c_{4} \chi^4\right) \dot{\chi}^2\right)
+24 \chi \left(1+2 c_{4} \chi^4\right) \left(5+6 c_{4} \chi^4\right) \ddot{\chi}\right)\right)\right\}\right],
\end{align}
%
where $\chi:=\dot{\phi}_0$.

\bibliography{draft_dhost_mcmc}

\begin{thebibliography}{54}%
\makeatletter
\providecommand \@ifxundefined [1]{%
 \@ifx{#1\undefined}
}%
\providecommand \@ifnum [1]{%
 \ifnum #1\expandafter \@firstoftwo
 \else \expandafter \@secondoftwo
 \fi
}%
\providecommand \@ifx [1]{%
 \ifx #1\expandafter \@firstoftwo
 \else \expandafter \@secondoftwo
 \fi
}%
\providecommand \natexlab [1]{#1}%
\providecommand \enquote  [1]{``#1''}%
\providecommand \bibnamefont  [1]{#1}%
\providecommand \bibfnamefont [1]{#1}%
\providecommand \citenamefont [1]{#1}%
\providecommand \href@noop [0]{\@secondoftwo}%
\providecommand \href [0]{\begingroup \@sanitize@url \@href}%
\providecommand \@href[1]{\@@startlink{#1}\@@href}%
\providecommand \@@href[1]{\endgroup#1\@@endlink}%
\providecommand \@sanitize@url [0]{\catcode `\\12\catcode `\$12\catcode
  `\&12\catcode `\#12\catcode `\^12\catcode `\_12\catcode `\%12\relax}%
\providecommand \@@startlink[1]{}%
\providecommand \@@endlink[0]{}%
\providecommand \url  [0]{\begingroup\@sanitize@url \@url }%
\providecommand \@url [1]{\endgroup\@href {#1}{\urlprefix }}%
\providecommand \urlprefix  [0]{URL }%
\providecommand \Eprint [0]{\href }%
\providecommand \doibase [0]{https://doi.org/}%
\providecommand \selectlanguage [0]{\@gobble}%
\providecommand \bibinfo  [0]{\@secondoftwo}%
\providecommand \bibfield  [0]{\@secondoftwo}%
\providecommand \translation [1]{[#1]}%
\providecommand \BibitemOpen [0]{}%
\providecommand \bibitemStop [0]{}%
\providecommand \bibitemNoStop [0]{.\EOS\space}%
\providecommand \EOS [0]{\spacefactor3000\relax}%
\providecommand \BibitemShut  [1]{\csname bibitem#1\endcsname}%
\let\auto@bib@innerbib\@empty
\bibitem [{\citenamefont {Horndeski}(1974)}]{Horndeski:1974wa}%
  \BibitemOpen
  \bibfield  {author} {\bibinfo {author} {\bibfnamefont {G.~W.}\ \bibnamefont
  {Horndeski}},\ }\href {https://doi.org/10.1007/BF01807638} {\bibfield
  {journal} {\bibinfo  {journal} {Int. J. Theor. Phys.}\ }\textbf {\bibinfo
  {volume} {10}},\ \bibinfo {pages} {363} (\bibinfo {year} {1974})}\BibitemShut
  {NoStop}%
\bibitem [{\citenamefont {Deffayet}\ \emph {et~al.}(2011)\citenamefont
  {Deffayet}, \citenamefont {Gao}, \citenamefont {Steer},\ and\ \citenamefont
  {Zahariade}}]{Deffayet:2011gz}%
  \BibitemOpen
  \bibfield  {author} {\bibinfo {author} {\bibfnamefont {C.}~\bibnamefont
  {Deffayet}}, \bibinfo {author} {\bibfnamefont {X.}~\bibnamefont {Gao}},
  \bibinfo {author} {\bibfnamefont {D.~A.}\ \bibnamefont {Steer}},\ and\
  \bibinfo {author} {\bibfnamefont {G.}~\bibnamefont {Zahariade}},\ }\href
  {https://doi.org/10.1103/PhysRevD.84.064039} {\bibfield  {journal} {\bibinfo
  {journal} {Phys. Rev. D}\ }\textbf {\bibinfo {volume} {84}},\ \bibinfo
  {pages} {064039} (\bibinfo {year} {2011})},\ \Eprint
  {https://arxiv.org/abs/1103.3260} {arXiv:1103.3260 [hep-th]} \BibitemShut
  {NoStop}%
\bibitem [{\citenamefont {Kobayashi}\ \emph {et~al.}(2011)\citenamefont
  {Kobayashi}, \citenamefont {Yamaguchi},\ and\ \citenamefont
  {Yokoyama}}]{Kobayashi:2011nu}%
  \BibitemOpen
  \bibfield  {author} {\bibinfo {author} {\bibfnamefont {T.}~\bibnamefont
  {Kobayashi}}, \bibinfo {author} {\bibfnamefont {M.}~\bibnamefont
  {Yamaguchi}},\ and\ \bibinfo {author} {\bibfnamefont {J.}~\bibnamefont
  {Yokoyama}},\ }\href {https://doi.org/10.1143/PTP.126.511} {\bibfield
  {journal} {\bibinfo  {journal} {Prog. Theor. Phys.}\ }\textbf {\bibinfo
  {volume} {126}},\ \bibinfo {pages} {511} (\bibinfo {year} {2011})},\ \Eprint
  {https://arxiv.org/abs/1105.5723} {arXiv:1105.5723 [hep-th]} \BibitemShut
  {NoStop}%
\bibitem [{\citenamefont {Kobayashi}(2019)}]{Kobayashi:2019hrl}%
  \BibitemOpen
  \bibfield  {author} {\bibinfo {author} {\bibfnamefont {T.}~\bibnamefont
  {Kobayashi}},\ }\href {https://doi.org/10.1088/1361-6633/ab2429} {\bibfield
  {journal} {\bibinfo  {journal} {Rept. Prog. Phys.}\ }\textbf {\bibinfo
  {volume} {82}},\ \bibinfo {pages} {086901} (\bibinfo {year} {2019})},\
  \Eprint {https://arxiv.org/abs/1901.07183} {arXiv:1901.07183 [gr-qc]}
  \BibitemShut {NoStop}%
\bibitem [{\citenamefont {Gleyzes}\ \emph
  {et~al.}(2015{\natexlab{a}})\citenamefont {Gleyzes}, \citenamefont
  {Langlois}, \citenamefont {Piazza},\ and\ \citenamefont
  {Vernizzi}}]{Gleyzes:2014dya}%
  \BibitemOpen
  \bibfield  {author} {\bibinfo {author} {\bibfnamefont {J.}~\bibnamefont
  {Gleyzes}}, \bibinfo {author} {\bibfnamefont {D.}~\bibnamefont {Langlois}},
  \bibinfo {author} {\bibfnamefont {F.}~\bibnamefont {Piazza}},\ and\ \bibinfo
  {author} {\bibfnamefont {F.}~\bibnamefont {Vernizzi}},\ }\href
  {https://doi.org/10.1103/PhysRevLett.114.211101} {\bibfield  {journal}
  {\bibinfo  {journal} {Phys. Rev. Lett.}\ }\textbf {\bibinfo {volume} {114}},\
  \bibinfo {pages} {211101} (\bibinfo {year} {2015}{\natexlab{a}})},\ \Eprint
  {https://arxiv.org/abs/1404.6495} {arXiv:1404.6495 [hep-th]} \BibitemShut
  {NoStop}%
\bibitem [{\citenamefont {Gleyzes}\ \emph
  {et~al.}(2015{\natexlab{b}})\citenamefont {Gleyzes}, \citenamefont
  {Langlois}, \citenamefont {Piazza},\ and\ \citenamefont
  {Vernizzi}}]{Gleyzes:2014qga}%
  \BibitemOpen
  \bibfield  {author} {\bibinfo {author} {\bibfnamefont {J.}~\bibnamefont
  {Gleyzes}}, \bibinfo {author} {\bibfnamefont {D.}~\bibnamefont {Langlois}},
  \bibinfo {author} {\bibfnamefont {F.}~\bibnamefont {Piazza}},\ and\ \bibinfo
  {author} {\bibfnamefont {F.}~\bibnamefont {Vernizzi}},\ }\href
  {https://doi.org/10.1088/1475-7516/2015/02/018} {\bibfield  {journal}
  {\bibinfo  {journal} {JCAP}\ }\textbf {\bibinfo {volume} {02}},\ \bibinfo
  {pages} {018}},\ \Eprint {https://arxiv.org/abs/1408.1952} {arXiv:1408.1952
  [astro-ph.CO]} \BibitemShut {NoStop}%
\bibitem [{\citenamefont {Langlois}\ and\ \citenamefont
  {Noui}(2016{\natexlab{a}})}]{Langlois:2015cwa}%
  \BibitemOpen
  \bibfield  {author} {\bibinfo {author} {\bibfnamefont {D.}~\bibnamefont
  {Langlois}}\ and\ \bibinfo {author} {\bibfnamefont {K.}~\bibnamefont
  {Noui}},\ }\href {https://doi.org/10.1088/1475-7516/2016/02/034} {\bibfield
  {journal} {\bibinfo  {journal} {JCAP}\ }\textbf {\bibinfo {volume} {02}},\
  \bibinfo {pages} {034}},\ \Eprint {https://arxiv.org/abs/1510.06930}
  {arXiv:1510.06930 [gr-qc]} \BibitemShut {NoStop}%
\bibitem [{\citenamefont {Langlois}\ and\ \citenamefont
  {Noui}(2016{\natexlab{b}})}]{Langlois:2015skt}%
  \BibitemOpen
  \bibfield  {author} {\bibinfo {author} {\bibfnamefont {D.}~\bibnamefont
  {Langlois}}\ and\ \bibinfo {author} {\bibfnamefont {K.}~\bibnamefont
  {Noui}},\ }\href {https://doi.org/10.1088/1475-7516/2016/07/016} {\bibfield
  {journal} {\bibinfo  {journal} {JCAP}\ }\textbf {\bibinfo {volume} {07}},\
  \bibinfo {pages} {016}},\ \Eprint {https://arxiv.org/abs/1512.06820}
  {arXiv:1512.06820 [gr-qc]} \BibitemShut {NoStop}%
\bibitem [{\citenamefont {Langlois}(2019)}]{Langlois:2018dxi}%
  \BibitemOpen
  \bibfield  {author} {\bibinfo {author} {\bibfnamefont {D.}~\bibnamefont
  {Langlois}},\ }\href {https://doi.org/10.1142/S0218271819420069} {\bibfield
  {journal} {\bibinfo  {journal} {Int. J. Mod. Phys. D}\ }\textbf {\bibinfo
  {volume} {28}},\ \bibinfo {pages} {1942006} (\bibinfo {year} {2019})},\
  \Eprint {https://arxiv.org/abs/1811.06271} {arXiv:1811.06271 [gr-qc]}
  \BibitemShut {NoStop}%
\bibitem [{\citenamefont {Ben~Achour}\ \emph {et~al.}(2016)\citenamefont
  {Ben~Achour}, \citenamefont {Langlois},\ and\ \citenamefont
  {Noui}}]{BenAchour:2016cay}%
  \BibitemOpen
  \bibfield  {author} {\bibinfo {author} {\bibfnamefont {J.}~\bibnamefont
  {Ben~Achour}}, \bibinfo {author} {\bibfnamefont {D.}~\bibnamefont
  {Langlois}},\ and\ \bibinfo {author} {\bibfnamefont {K.}~\bibnamefont
  {Noui}},\ }\href {https://doi.org/10.1103/PhysRevD.93.124005} {\bibfield
  {journal} {\bibinfo  {journal} {Phys. Rev. D}\ }\textbf {\bibinfo {volume}
  {93}},\ \bibinfo {pages} {124005} (\bibinfo {year} {2016})},\ \Eprint
  {https://arxiv.org/abs/1602.08398} {arXiv:1602.08398 [gr-qc]} \BibitemShut
  {NoStop}%
\bibitem [{\citenamefont {Crisostomi}\ \emph {et~al.}(2016)\citenamefont
  {Crisostomi}, \citenamefont {Koyama},\ and\ \citenamefont
  {Tasinato}}]{Crisostomi:2016czh}%
  \BibitemOpen
  \bibfield  {author} {\bibinfo {author} {\bibfnamefont {M.}~\bibnamefont
  {Crisostomi}}, \bibinfo {author} {\bibfnamefont {K.}~\bibnamefont {Koyama}},\
  and\ \bibinfo {author} {\bibfnamefont {G.}~\bibnamefont {Tasinato}},\ }\href
  {https://doi.org/10.1088/1475-7516/2016/04/044} {\bibfield  {journal}
  {\bibinfo  {journal} {JCAP}\ }\textbf {\bibinfo {volume} {04}},\ \bibinfo
  {pages} {044}},\ \Eprint {https://arxiv.org/abs/1602.03119} {arXiv:1602.03119
  [hep-th]} \BibitemShut {NoStop}%
\bibitem [{\citenamefont {De~Felice}\ \emph {et~al.}(2015)\citenamefont
  {De~Felice}, \citenamefont {Koyama},\ and\ \citenamefont
  {Tsujikawa}}]{DeFelice:2015isa}%
  \BibitemOpen
  \bibfield  {author} {\bibinfo {author} {\bibfnamefont {A.}~\bibnamefont
  {De~Felice}}, \bibinfo {author} {\bibfnamefont {K.}~\bibnamefont {Koyama}},\
  and\ \bibinfo {author} {\bibfnamefont {S.}~\bibnamefont {Tsujikawa}},\ }\href
  {https://doi.org/10.1088/1475-7516/2015/05/058} {\bibfield  {journal}
  {\bibinfo  {journal} {JCAP}\ }\textbf {\bibinfo {volume} {05}},\ \bibinfo
  {pages} {058}},\ \Eprint {https://arxiv.org/abs/1503.06539} {arXiv:1503.06539
  [gr-qc]} \BibitemShut {NoStop}%
\bibitem [{\citenamefont {D'Amico}\ \emph {et~al.}(2017)\citenamefont
  {D'Amico}, \citenamefont {Huang}, \citenamefont {Mancarella},\ and\
  \citenamefont {Vernizzi}}]{DAmico:2016ntq}%
  \BibitemOpen
  \bibfield  {author} {\bibinfo {author} {\bibfnamefont {G.}~\bibnamefont
  {D'Amico}}, \bibinfo {author} {\bibfnamefont {Z.}~\bibnamefont {Huang}},
  \bibinfo {author} {\bibfnamefont {M.}~\bibnamefont {Mancarella}},\ and\
  \bibinfo {author} {\bibfnamefont {F.}~\bibnamefont {Vernizzi}},\ }\href
  {https://doi.org/10.1088/1475-7516/2017/02/014} {\bibfield  {journal}
  {\bibinfo  {journal} {JCAP}\ }\textbf {\bibinfo {volume} {02}},\ \bibinfo
  {pages} {014}},\ \Eprint {https://arxiv.org/abs/1609.01272} {arXiv:1609.01272
  [astro-ph.CO]} \BibitemShut {NoStop}%
\bibitem [{\citenamefont {Renk}\ \emph {et~al.}(2016)\citenamefont {Renk},
  \citenamefont {Zumalacarregui},\ and\ \citenamefont
  {Montanari}}]{Renk:2016olm}%
  \BibitemOpen
  \bibfield  {author} {\bibinfo {author} {\bibfnamefont {J.}~\bibnamefont
  {Renk}}, \bibinfo {author} {\bibfnamefont {M.}~\bibnamefont
  {Zumalacarregui}},\ and\ \bibinfo {author} {\bibfnamefont {F.}~\bibnamefont
  {Montanari}},\ }\href {https://doi.org/10.1088/1475-7516/2016/07/040}
  {\bibfield  {journal} {\bibinfo  {journal} {JCAP}\ }\textbf {\bibinfo
  {volume} {07}},\ \bibinfo {pages} {040}},\ \Eprint
  {https://arxiv.org/abs/1604.03487} {arXiv:1604.03487 [astro-ph.CO]}
  \BibitemShut {NoStop}%
\bibitem [{\citenamefont {Kreisch}\ and\ \citenamefont
  {Komatsu}(2018)}]{Kreisch:2017uet}%
  \BibitemOpen
  \bibfield  {author} {\bibinfo {author} {\bibfnamefont {C.~D.}\ \bibnamefont
  {Kreisch}}\ and\ \bibinfo {author} {\bibfnamefont {E.}~\bibnamefont
  {Komatsu}},\ }\href {https://doi.org/10.1088/1475-7516/2018/12/030}
  {\bibfield  {journal} {\bibinfo  {journal} {JCAP}\ }\textbf {\bibinfo
  {volume} {12}},\ \bibinfo {pages} {030}},\ \Eprint
  {https://arxiv.org/abs/1712.02710} {arXiv:1712.02710 [astro-ph.CO]}
  \BibitemShut {NoStop}%
\bibitem [{\citenamefont {Namikawa}\ \emph {et~al.}(2018)\citenamefont
  {Namikawa}, \citenamefont {Bouchet},\ and\ \citenamefont
  {Taruya}}]{Namikawa:2018erh}%
  \BibitemOpen
  \bibfield  {author} {\bibinfo {author} {\bibfnamefont {T.}~\bibnamefont
  {Namikawa}}, \bibinfo {author} {\bibfnamefont {F.~R.}\ \bibnamefont
  {Bouchet}},\ and\ \bibinfo {author} {\bibfnamefont {A.}~\bibnamefont
  {Taruya}},\ }\href {https://doi.org/10.1103/PhysRevD.98.043530} {\bibfield
  {journal} {\bibinfo  {journal} {Phys. Rev. D}\ }\textbf {\bibinfo {volume}
  {98}},\ \bibinfo {pages} {043530} (\bibinfo {year} {2018})},\ \Eprint
  {https://arxiv.org/abs/1805.10567} {arXiv:1805.10567 [astro-ph.CO]}
  \BibitemShut {NoStop}%
\bibitem [{\citenamefont {Noller}\ and\ \citenamefont
  {Nicola}(2019)}]{Noller:2018wyv}%
  \BibitemOpen
  \bibfield  {author} {\bibinfo {author} {\bibfnamefont {J.}~\bibnamefont
  {Noller}}\ and\ \bibinfo {author} {\bibfnamefont {A.}~\bibnamefont
  {Nicola}},\ }\href {https://doi.org/10.1103/PhysRevD.99.103502} {\bibfield
  {journal} {\bibinfo  {journal} {Phys. Rev. D}\ }\textbf {\bibinfo {volume}
  {99}},\ \bibinfo {pages} {103502} (\bibinfo {year} {2019})},\ \Eprint
  {https://arxiv.org/abs/1811.12928} {arXiv:1811.12928 [astro-ph.CO]}
  \BibitemShut {NoStop}%
\bibitem [{\citenamefont {Peirone}\ \emph {et~al.}(2019)\citenamefont
  {Peirone}, \citenamefont {Benevento}, \citenamefont {Frusciante},\ and\
  \citenamefont {Tsujikawa}}]{Peirone:2019yjs}%
  \BibitemOpen
  \bibfield  {author} {\bibinfo {author} {\bibfnamefont {S.}~\bibnamefont
  {Peirone}}, \bibinfo {author} {\bibfnamefont {G.}~\bibnamefont {Benevento}},
  \bibinfo {author} {\bibfnamefont {N.}~\bibnamefont {Frusciante}},\ and\
  \bibinfo {author} {\bibfnamefont {S.}~\bibnamefont {Tsujikawa}},\ }\href
  {https://doi.org/10.1103/PhysRevD.100.063509} {\bibfield  {journal} {\bibinfo
   {journal} {Phys. Rev. D}\ }\textbf {\bibinfo {volume} {100}},\ \bibinfo
  {pages} {063509} (\bibinfo {year} {2019})},\ \Eprint
  {https://arxiv.org/abs/1905.11364} {arXiv:1905.11364 [astro-ph.CO]}
  \BibitemShut {NoStop}%
\bibitem [{\citenamefont {Traykova}\ \emph {et~al.}(2019)\citenamefont
  {Traykova}, \citenamefont {Bellini},\ and\ \citenamefont
  {Ferreira}}]{Traykova:2019oyx}%
  \BibitemOpen
  \bibfield  {author} {\bibinfo {author} {\bibfnamefont {D.}~\bibnamefont
  {Traykova}}, \bibinfo {author} {\bibfnamefont {E.}~\bibnamefont {Bellini}},\
  and\ \bibinfo {author} {\bibfnamefont {P.~G.}\ \bibnamefont {Ferreira}},\
  }\href {https://doi.org/10.1088/1475-7516/2019/08/035} {\bibfield  {journal}
  {\bibinfo  {journal} {JCAP}\ }\textbf {\bibinfo {volume} {08}},\ \bibinfo
  {pages} {035}},\ \Eprint {https://arxiv.org/abs/1902.10687} {arXiv:1902.10687
  [astro-ph.CO]} \BibitemShut {NoStop}%
\bibitem [{\citenamefont {Sakstein}(2015{\natexlab{a}})}]{Sakstein:2015zoa}%
  \BibitemOpen
  \bibfield  {author} {\bibinfo {author} {\bibfnamefont {J.}~\bibnamefont
  {Sakstein}},\ }\href {https://doi.org/10.1103/PhysRevLett.115.201101}
  {\bibfield  {journal} {\bibinfo  {journal} {Phys. Rev. Lett.}\ }\textbf
  {\bibinfo {volume} {115}},\ \bibinfo {pages} {201101} (\bibinfo {year}
  {2015}{\natexlab{a}})},\ \Eprint {https://arxiv.org/abs/1510.05964}
  {arXiv:1510.05964 [astro-ph.CO]} \BibitemShut {NoStop}%
\bibitem [{\citenamefont {Sakstein}(2015{\natexlab{b}})}]{Sakstein:2015aac}%
  \BibitemOpen
  \bibfield  {author} {\bibinfo {author} {\bibfnamefont {J.}~\bibnamefont
  {Sakstein}},\ }\href {https://doi.org/10.1103/PhysRevD.92.124045} {\bibfield
  {journal} {\bibinfo  {journal} {Phys. Rev. D}\ }\textbf {\bibinfo {volume}
  {92}},\ \bibinfo {pages} {124045} (\bibinfo {year} {2015}{\natexlab{b}})},\
  \Eprint {https://arxiv.org/abs/1511.01685} {arXiv:1511.01685 [astro-ph.CO]}
  \BibitemShut {NoStop}%
\bibitem [{\citenamefont {Jain}\ \emph {et~al.}(2016)\citenamefont {Jain},
  \citenamefont {Kouvaris},\ and\ \citenamefont {Nielsen}}]{Jain:2015edg}%
  \BibitemOpen
  \bibfield  {author} {\bibinfo {author} {\bibfnamefont {R.~K.}\ \bibnamefont
  {Jain}}, \bibinfo {author} {\bibfnamefont {C.}~\bibnamefont {Kouvaris}},\
  and\ \bibinfo {author} {\bibfnamefont {N.~G.}\ \bibnamefont {Nielsen}},\
  }\href {https://doi.org/10.1103/PhysRevLett.116.151103} {\bibfield  {journal}
  {\bibinfo  {journal} {Phys. Rev. Lett.}\ }\textbf {\bibinfo {volume} {116}},\
  \bibinfo {pages} {151103} (\bibinfo {year} {2016})},\ \Eprint
  {https://arxiv.org/abs/1512.05946} {arXiv:1512.05946 [astro-ph.CO]}
  \BibitemShut {NoStop}%
\bibitem [{\citenamefont {Sakstein}\ \emph {et~al.}(2016)\citenamefont
  {Sakstein}, \citenamefont {Wilcox}, \citenamefont {Bacon}, \citenamefont
  {Koyama},\ and\ \citenamefont {Nichol}}]{Sakstein:2016ggl}%
  \BibitemOpen
  \bibfield  {author} {\bibinfo {author} {\bibfnamefont {J.}~\bibnamefont
  {Sakstein}}, \bibinfo {author} {\bibfnamefont {H.}~\bibnamefont {Wilcox}},
  \bibinfo {author} {\bibfnamefont {D.}~\bibnamefont {Bacon}}, \bibinfo
  {author} {\bibfnamefont {K.}~\bibnamefont {Koyama}},\ and\ \bibinfo {author}
  {\bibfnamefont {R.~C.}\ \bibnamefont {Nichol}},\ }\href
  {https://doi.org/10.1088/1475-7516/2016/07/019} {\bibfield  {journal}
  {\bibinfo  {journal} {JCAP}\ }\textbf {\bibinfo {volume} {07}},\ \bibinfo
  {pages} {019}},\ \Eprint {https://arxiv.org/abs/1603.06368} {arXiv:1603.06368
  [astro-ph.CO]} \BibitemShut {NoStop}%
\bibitem [{\citenamefont {Sakstein}\ \emph
  {et~al.}(2017{\natexlab{a}})\citenamefont {Sakstein}, \citenamefont
  {Kenna-Allison},\ and\ \citenamefont {Koyama}}]{Sakstein:2016lyj}%
  \BibitemOpen
  \bibfield  {author} {\bibinfo {author} {\bibfnamefont {J.}~\bibnamefont
  {Sakstein}}, \bibinfo {author} {\bibfnamefont {M.}~\bibnamefont
  {Kenna-Allison}},\ and\ \bibinfo {author} {\bibfnamefont {K.}~\bibnamefont
  {Koyama}},\ }\href {https://doi.org/10.1088/1475-7516/2017/03/007} {\bibfield
   {journal} {\bibinfo  {journal} {JCAP}\ }\textbf {\bibinfo {volume} {03}},\
  \bibinfo {pages} {007}},\ \Eprint {https://arxiv.org/abs/1611.01062}
  {arXiv:1611.01062 [gr-qc]} \BibitemShut {NoStop}%
\bibitem [{\citenamefont {Salzano}\ \emph {et~al.}(2017)\citenamefont
  {Salzano}, \citenamefont {Mota}, \citenamefont {Capozziello},\ and\
  \citenamefont {Donahue}}]{Salzano:2017qac}%
  \BibitemOpen
  \bibfield  {author} {\bibinfo {author} {\bibfnamefont {V.}~\bibnamefont
  {Salzano}}, \bibinfo {author} {\bibfnamefont {D.~F.}\ \bibnamefont {Mota}},
  \bibinfo {author} {\bibfnamefont {S.}~\bibnamefont {Capozziello}},\ and\
  \bibinfo {author} {\bibfnamefont {M.}~\bibnamefont {Donahue}},\ }\href
  {https://doi.org/10.1103/PhysRevD.95.044038} {\bibfield  {journal} {\bibinfo
  {journal} {Phys. Rev. D}\ }\textbf {\bibinfo {volume} {95}},\ \bibinfo
  {pages} {044038} (\bibinfo {year} {2017})},\ \Eprint
  {https://arxiv.org/abs/1701.03517} {arXiv:1701.03517 [astro-ph.CO]}
  \BibitemShut {NoStop}%
\bibitem [{\citenamefont {Babichev}\ \emph {et~al.}(2016)\citenamefont
  {Babichev}, \citenamefont {Koyama}, \citenamefont {Langlois}, \citenamefont
  {Saito},\ and\ \citenamefont {Sakstein}}]{Babichev:2016jom}%
  \BibitemOpen
  \bibfield  {author} {\bibinfo {author} {\bibfnamefont {E.}~\bibnamefont
  {Babichev}}, \bibinfo {author} {\bibfnamefont {K.}~\bibnamefont {Koyama}},
  \bibinfo {author} {\bibfnamefont {D.}~\bibnamefont {Langlois}}, \bibinfo
  {author} {\bibfnamefont {R.}~\bibnamefont {Saito}},\ and\ \bibinfo {author}
  {\bibfnamefont {J.}~\bibnamefont {Sakstein}},\ }\href
  {https://doi.org/10.1088/0264-9381/33/23/235014} {\bibfield  {journal}
  {\bibinfo  {journal} {Class. Quant. Grav.}\ }\textbf {\bibinfo {volume}
  {33}},\ \bibinfo {pages} {235014} (\bibinfo {year} {2016})},\ \Eprint
  {https://arxiv.org/abs/1606.06627} {arXiv:1606.06627 [gr-qc]} \BibitemShut
  {NoStop}%
\bibitem [{\citenamefont {Sakstein}\ \emph
  {et~al.}(2017{\natexlab{b}})\citenamefont {Sakstein}, \citenamefont
  {Babichev}, \citenamefont {Koyama}, \citenamefont {Langlois},\ and\
  \citenamefont {Saito}}]{Sakstein:2016oel}%
  \BibitemOpen
  \bibfield  {author} {\bibinfo {author} {\bibfnamefont {J.}~\bibnamefont
  {Sakstein}}, \bibinfo {author} {\bibfnamefont {E.}~\bibnamefont {Babichev}},
  \bibinfo {author} {\bibfnamefont {K.}~\bibnamefont {Koyama}}, \bibinfo
  {author} {\bibfnamefont {D.}~\bibnamefont {Langlois}},\ and\ \bibinfo
  {author} {\bibfnamefont {R.}~\bibnamefont {Saito}},\ }\href
  {https://doi.org/10.1103/PhysRevD.95.064013} {\bibfield  {journal} {\bibinfo
  {journal} {Phys. Rev. D}\ }\textbf {\bibinfo {volume} {95}},\ \bibinfo
  {pages} {064013} (\bibinfo {year} {2017}{\natexlab{b}})},\ \Eprint
  {https://arxiv.org/abs/1612.04263} {arXiv:1612.04263 [gr-qc]} \BibitemShut
  {NoStop}%
\bibitem [{\citenamefont {Hiramatsu}\ and\ \citenamefont
  {Yamauchi}(2020)}]{Hiramatsu:2020fcd}%
  \BibitemOpen
  \bibfield  {author} {\bibinfo {author} {\bibfnamefont {T.}~\bibnamefont
  {Hiramatsu}}\ and\ \bibinfo {author} {\bibfnamefont {D.}~\bibnamefont
  {Yamauchi}},\ }\href {https://doi.org/10.1103/PhysRevD.102.083525} {\bibfield
   {journal} {\bibinfo  {journal} {Phys. Rev. D}\ }\textbf {\bibinfo {volume}
  {102}},\ \bibinfo {pages} {083525} (\bibinfo {year} {2020})},\ \Eprint
  {https://arxiv.org/abs/2004.09520} {arXiv:2004.09520 [astro-ph.CO]}
  \BibitemShut {NoStop}%
\bibitem [{\citenamefont {Aghanim}\ \emph
  {et~al.}(2020{\natexlab{a}})\citenamefont {Aghanim} \emph
  {et~al.}}]{Planck:2018vyg}%
  \BibitemOpen
  \bibfield  {author} {\bibinfo {author} {\bibfnamefont {N.}~\bibnamefont
  {Aghanim}} \emph {et~al.} (\bibinfo {collaboration} {Planck}),\ }\href
  {https://doi.org/10.1051/0004-6361/201833910} {\bibfield  {journal} {\bibinfo
   {journal} {Astron. Astrophys.}\ }\textbf {\bibinfo {volume} {641}},\
  \bibinfo {pages} {A6} (\bibinfo {year} {2020}{\natexlab{a}})},\ \bibinfo
  {note} {[Erratum: Astron.Astrophys. 652, C4 (2021)]},\ \Eprint
  {https://arxiv.org/abs/1807.06209} {arXiv:1807.06209 [astro-ph.CO]}
  \BibitemShut {NoStop}%
\bibitem [{\citenamefont {Kobayashi}\ and\ \citenamefont
  {Hiramatsu}(2018)}]{Kobayashi:2018xvr}%
  \BibitemOpen
  \bibfield  {author} {\bibinfo {author} {\bibfnamefont {T.}~\bibnamefont
  {Kobayashi}}\ and\ \bibinfo {author} {\bibfnamefont {T.}~\bibnamefont
  {Hiramatsu}},\ }\href {https://doi.org/10.1103/PhysRevD.97.104012} {\bibfield
   {journal} {\bibinfo  {journal} {Phys. Rev. D}\ }\textbf {\bibinfo {volume}
  {97}},\ \bibinfo {pages} {104012} (\bibinfo {year} {2018})},\ \Eprint
  {https://arxiv.org/abs/1803.10510} {arXiv:1803.10510 [gr-qc]} \BibitemShut
  {NoStop}%
\bibitem [{\citenamefont {Chagoya}\ and\ \citenamefont
  {Tasinato}(2018)}]{Chagoya:2018lmv}%
  \BibitemOpen
  \bibfield  {author} {\bibinfo {author} {\bibfnamefont {J.}~\bibnamefont
  {Chagoya}}\ and\ \bibinfo {author} {\bibfnamefont {G.}~\bibnamefont
  {Tasinato}},\ }\href {https://doi.org/10.1088/1475-7516/2018/08/006}
  {\bibfield  {journal} {\bibinfo  {journal} {JCAP}\ }\textbf {\bibinfo
  {volume} {08}},\ \bibinfo {pages} {006}},\ \Eprint
  {https://arxiv.org/abs/1803.07476} {arXiv:1803.07476 [gr-qc]} \BibitemShut
  {NoStop}%
\bibitem [{\citenamefont {Gubitosi}\ \emph {et~al.}(2013)\citenamefont
  {Gubitosi}, \citenamefont {Piazza},\ and\ \citenamefont
  {Vernizzi}}]{Gubitosi:2012hu}%
  \BibitemOpen
  \bibfield  {author} {\bibinfo {author} {\bibfnamefont {G.}~\bibnamefont
  {Gubitosi}}, \bibinfo {author} {\bibfnamefont {F.}~\bibnamefont {Piazza}},\
  and\ \bibinfo {author} {\bibfnamefont {F.}~\bibnamefont {Vernizzi}},\ }\href
  {https://doi.org/10.1088/1475-7516/2013/02/032} {\bibfield  {journal}
  {\bibinfo  {journal} {JCAP}\ }\textbf {\bibinfo {volume} {02}},\ \bibinfo
  {pages} {032}},\ \Eprint {https://arxiv.org/abs/1210.0201} {arXiv:1210.0201
  [hep-th]} \BibitemShut {NoStop}%
\bibitem [{\citenamefont {Bloomfield}\ \emph {et~al.}(2013)\citenamefont
  {Bloomfield}, \citenamefont {Flanagan}, \citenamefont {Park},\ and\
  \citenamefont {Watson}}]{Bloomfield:2012ff}%
  \BibitemOpen
  \bibfield  {author} {\bibinfo {author} {\bibfnamefont {J.~K.}\ \bibnamefont
  {Bloomfield}}, \bibinfo {author} {\bibfnamefont {E.~E.}\ \bibnamefont
  {Flanagan}}, \bibinfo {author} {\bibfnamefont {M.}~\bibnamefont {Park}},\
  and\ \bibinfo {author} {\bibfnamefont {S.}~\bibnamefont {Watson}},\ }\href
  {https://doi.org/10.1088/1475-7516/2013/08/010} {\bibfield  {journal}
  {\bibinfo  {journal} {JCAP}\ }\textbf {\bibinfo {volume} {08}},\ \bibinfo
  {pages} {010}},\ \Eprint {https://arxiv.org/abs/1211.7054} {arXiv:1211.7054
  [astro-ph.CO]} \BibitemShut {NoStop}%
\bibitem [{\citenamefont {Gleyzes}\ \emph {et~al.}(2013)\citenamefont
  {Gleyzes}, \citenamefont {Langlois}, \citenamefont {Piazza},\ and\
  \citenamefont {Vernizzi}}]{Gleyzes:2013ooa}%
  \BibitemOpen
  \bibfield  {author} {\bibinfo {author} {\bibfnamefont {J.}~\bibnamefont
  {Gleyzes}}, \bibinfo {author} {\bibfnamefont {D.}~\bibnamefont {Langlois}},
  \bibinfo {author} {\bibfnamefont {F.}~\bibnamefont {Piazza}},\ and\ \bibinfo
  {author} {\bibfnamefont {F.}~\bibnamefont {Vernizzi}},\ }\href
  {https://doi.org/10.1088/1475-7516/2013/08/025} {\bibfield  {journal}
  {\bibinfo  {journal} {JCAP}\ }\textbf {\bibinfo {volume} {08}},\ \bibinfo
  {pages} {025}},\ \Eprint {https://arxiv.org/abs/1304.4840} {arXiv:1304.4840
  [hep-th]} \BibitemShut {NoStop}%
\bibitem [{\citenamefont {Bloomfield}(2013)}]{Bloomfield:2013efa}%
  \BibitemOpen
  \bibfield  {author} {\bibinfo {author} {\bibfnamefont {J.}~\bibnamefont
  {Bloomfield}},\ }\href {https://doi.org/10.1088/1475-7516/2013/12/044}
  {\bibfield  {journal} {\bibinfo  {journal} {JCAP}\ }\textbf {\bibinfo
  {volume} {12}},\ \bibinfo {pages} {044}},\ \Eprint
  {https://arxiv.org/abs/1304.6712} {arXiv:1304.6712 [astro-ph.CO]}
  \BibitemShut {NoStop}%
\bibitem [{\citenamefont {Piazza}\ and\ \citenamefont
  {Vernizzi}(2013)}]{Piazza:2013coa}%
  \BibitemOpen
  \bibfield  {author} {\bibinfo {author} {\bibfnamefont {F.}~\bibnamefont
  {Piazza}}\ and\ \bibinfo {author} {\bibfnamefont {F.}~\bibnamefont
  {Vernizzi}},\ }\href {https://doi.org/10.1088/0264-9381/30/21/214007}
  {\bibfield  {journal} {\bibinfo  {journal} {Class. Quant. Grav.}\ }\textbf
  {\bibinfo {volume} {30}},\ \bibinfo {pages} {214007} (\bibinfo {year}
  {2013})},\ \Eprint {https://arxiv.org/abs/1307.4350} {arXiv:1307.4350
  [hep-th]} \BibitemShut {NoStop}%
\bibitem [{\citenamefont {Gleyzes}\ \emph
  {et~al.}(2015{\natexlab{c}})\citenamefont {Gleyzes}, \citenamefont
  {Langlois},\ and\ \citenamefont {Vernizzi}}]{Gleyzes:2014rba}%
  \BibitemOpen
  \bibfield  {author} {\bibinfo {author} {\bibfnamefont {J.}~\bibnamefont
  {Gleyzes}}, \bibinfo {author} {\bibfnamefont {D.}~\bibnamefont {Langlois}},\
  and\ \bibinfo {author} {\bibfnamefont {F.}~\bibnamefont {Vernizzi}},\ }\href
  {https://doi.org/10.1142/S021827181443010X} {\bibfield  {journal} {\bibinfo
  {journal} {Int. J. Mod. Phys. D}\ }\textbf {\bibinfo {volume} {23}},\
  \bibinfo {pages} {1443010} (\bibinfo {year} {2015}{\natexlab{c}})},\ \Eprint
  {https://arxiv.org/abs/1411.3712} {arXiv:1411.3712 [hep-th]} \BibitemShut
  {NoStop}%
\bibitem [{\citenamefont {Gleyzes}\ \emph
  {et~al.}(2015{\natexlab{d}})\citenamefont {Gleyzes}, \citenamefont
  {Langlois}, \citenamefont {Mancarella},\ and\ \citenamefont
  {Vernizzi}}]{Gleyzes:2015pma}%
  \BibitemOpen
  \bibfield  {author} {\bibinfo {author} {\bibfnamefont {J.}~\bibnamefont
  {Gleyzes}}, \bibinfo {author} {\bibfnamefont {D.}~\bibnamefont {Langlois}},
  \bibinfo {author} {\bibfnamefont {M.}~\bibnamefont {Mancarella}},\ and\
  \bibinfo {author} {\bibfnamefont {F.}~\bibnamefont {Vernizzi}},\ }\href
  {https://doi.org/10.1088/1475-7516/2015/08/054} {\bibfield  {journal}
  {\bibinfo  {journal} {JCAP}\ }\textbf {\bibinfo {volume} {08}},\ \bibinfo
  {pages} {054}},\ \Eprint {https://arxiv.org/abs/1504.05481} {arXiv:1504.05481
  [astro-ph.CO]} \BibitemShut {NoStop}%
\bibitem [{\citenamefont {Langlois}\ \emph {et~al.}(2017)\citenamefont
  {Langlois}, \citenamefont {Mancarella}, \citenamefont {Noui},\ and\
  \citenamefont {Vernizzi}}]{Langlois:2017mxy}%
  \BibitemOpen
  \bibfield  {author} {\bibinfo {author} {\bibfnamefont {D.}~\bibnamefont
  {Langlois}}, \bibinfo {author} {\bibfnamefont {M.}~\bibnamefont
  {Mancarella}}, \bibinfo {author} {\bibfnamefont {K.}~\bibnamefont {Noui}},\
  and\ \bibinfo {author} {\bibfnamefont {F.}~\bibnamefont {Vernizzi}},\ }\href
  {https://doi.org/10.1088/1475-7516/2017/05/033} {\bibfield  {journal}
  {\bibinfo  {journal} {JCAP}\ }\textbf {\bibinfo {volume} {05}},\ \bibinfo
  {pages} {033}},\ \Eprint {https://arxiv.org/abs/1703.03797} {arXiv:1703.03797
  [hep-th]} \BibitemShut {NoStop}%
\bibitem [{\citenamefont {Crisostomi}\ and\ \citenamefont
  {Koyama}(2018)}]{Crisostomi:2017pjs}%
  \BibitemOpen
  \bibfield  {author} {\bibinfo {author} {\bibfnamefont {M.}~\bibnamefont
  {Crisostomi}}\ and\ \bibinfo {author} {\bibfnamefont {K.}~\bibnamefont
  {Koyama}},\ }\href {https://doi.org/10.1103/PhysRevD.97.084004} {\bibfield
  {journal} {\bibinfo  {journal} {Phys. Rev. D}\ }\textbf {\bibinfo {volume}
  {97}},\ \bibinfo {pages} {084004} (\bibinfo {year} {2018})},\ \Eprint
  {https://arxiv.org/abs/1712.06556} {arXiv:1712.06556 [astro-ph.CO]}
  \BibitemShut {NoStop}%
\bibitem [{\citenamefont {Bellini}\ and\ \citenamefont
  {Sawicki}(2014)}]{Bellini:2014fua}%
  \BibitemOpen
  \bibfield  {author} {\bibinfo {author} {\bibfnamefont {E.}~\bibnamefont
  {Bellini}}\ and\ \bibinfo {author} {\bibfnamefont {I.}~\bibnamefont
  {Sawicki}},\ }\href {https://doi.org/10.1088/1475-7516/2014/07/050}
  {\bibfield  {journal} {\bibinfo  {journal} {JCAP}\ }\textbf {\bibinfo
  {volume} {07}},\ \bibinfo {pages} {050}},\ \Eprint
  {https://arxiv.org/abs/1404.3713} {arXiv:1404.3713 [astro-ph.CO]}
  \BibitemShut {NoStop}%
\bibitem [{\citenamefont {Abbott}\ \emph
  {et~al.}(2017{\natexlab{a}})\citenamefont {Abbott} \emph
  {et~al.}}]{LIGOScientific:2017vwq}%
  \BibitemOpen
  \bibfield  {author} {\bibinfo {author} {\bibfnamefont {B.~P.}\ \bibnamefont
  {Abbott}} \emph {et~al.} (\bibinfo {collaboration} {LIGO Scientific,
  Virgo}),\ }\href {https://doi.org/10.1103/PhysRevLett.119.161101} {\bibfield
  {journal} {\bibinfo  {journal} {Phys. Rev. Lett.}\ }\textbf {\bibinfo
  {volume} {119}},\ \bibinfo {pages} {161101} (\bibinfo {year}
  {2017}{\natexlab{a}})},\ \Eprint {https://arxiv.org/abs/1710.05832}
  {arXiv:1710.05832 [gr-qc]} \BibitemShut {NoStop}%
\bibitem [{\citenamefont {Abbott}\ \emph
  {et~al.}(2017{\natexlab{b}})\citenamefont {Abbott} \emph
  {et~al.}}]{LIGOScientific:2017zic}%
  \BibitemOpen
  \bibfield  {author} {\bibinfo {author} {\bibfnamefont {B.~P.}\ \bibnamefont
  {Abbott}} \emph {et~al.} (\bibinfo {collaboration} {LIGO Scientific, Virgo,
  Fermi-GBM, INTEGRAL}),\ }\href {https://doi.org/10.3847/2041-8213/aa920c}
  {\bibfield  {journal} {\bibinfo  {journal} {Astrophys. J. Lett.}\ }\textbf
  {\bibinfo {volume} {848}},\ \bibinfo {pages} {L13} (\bibinfo {year}
  {2017}{\natexlab{b}})},\ \Eprint {https://arxiv.org/abs/1710.05834}
  {arXiv:1710.05834 [astro-ph.HE]} \BibitemShut {NoStop}%
\bibitem [{\citenamefont {Abbott}\ \emph
  {et~al.}(2017{\natexlab{c}})\citenamefont {Abbott} \emph
  {et~al.}}]{LIGOScientific:2017ync}%
  \BibitemOpen
  \bibfield  {author} {\bibinfo {author} {\bibfnamefont {B.~P.}\ \bibnamefont
  {Abbott}} \emph {et~al.} (\bibinfo {collaboration} {LIGO Scientific, Virgo,
  Fermi GBM, INTEGRAL, IceCube, AstroSat Cadmium Zinc Telluride Imager Team,
  IPN, Insight-Hxmt, ANTARES, Swift, AGILE Team, 1M2H Team, Dark Energy Camera
  GW-EM, DES, DLT40, GRAWITA, Fermi-LAT, ATCA, ASKAP, Las Cumbres Observatory
  Group, OzGrav, DWF (Deeper Wider Faster Program), AST3, CAASTRO, VINROUGE,
  MASTER, J-GEM, GROWTH, JAGWAR, CaltechNRAO, TTU-NRAO, NuSTAR, Pan-STARRS,
  MAXI Team, TZAC Consortium, KU, Nordic Optical Telescope, ePESSTO, GROND,
  Texas Tech University, SALT Group, TOROS, BOOTES, MWA, CALET, IKI-GW
  Follow-up, H.E.S.S., LOFAR, LWA, HAWC, Pierre Auger, ALMA, Euro VLBI Team, Pi
  of Sky, Chandra Team at McGill University, DFN, ATLAS Telescopes, High Time
  Resolution Universe Survey, RIMAS, RATIR, SKA South Africa/MeerKAT}),\ }\href
  {https://doi.org/10.3847/2041-8213/aa91c9} {\bibfield  {journal} {\bibinfo
  {journal} {Astrophys. J. Lett.}\ }\textbf {\bibinfo {volume} {848}},\
  \bibinfo {pages} {L12} (\bibinfo {year} {2017}{\natexlab{c}})},\ \Eprint
  {https://arxiv.org/abs/1710.05833} {arXiv:1710.05833 [astro-ph.HE]}
  \BibitemShut {NoStop}%
\bibitem [{\citenamefont {Creminelli}\ \emph {et~al.}(2018)\citenamefont
  {Creminelli}, \citenamefont {Lewandowski}, \citenamefont {Tambalo},\ and\
  \citenamefont {Vernizzi}}]{Creminelli:2018xsv}%
  \BibitemOpen
  \bibfield  {author} {\bibinfo {author} {\bibfnamefont {P.}~\bibnamefont
  {Creminelli}}, \bibinfo {author} {\bibfnamefont {M.}~\bibnamefont
  {Lewandowski}}, \bibinfo {author} {\bibfnamefont {G.}~\bibnamefont
  {Tambalo}},\ and\ \bibinfo {author} {\bibfnamefont {F.}~\bibnamefont
  {Vernizzi}},\ }\href {https://doi.org/10.1088/1475-7516/2018/12/025}
  {\bibfield  {journal} {\bibinfo  {journal} {JCAP}\ }\textbf {\bibinfo
  {volume} {12}},\ \bibinfo {pages} {025}},\ \Eprint
  {https://arxiv.org/abs/1809.03484} {arXiv:1809.03484 [astro-ph.CO]}
  \BibitemShut {NoStop}%
\bibitem [{\citenamefont {de~Rham}\ and\ \citenamefont
  {Melville}(2018)}]{deRham:2018red}%
  \BibitemOpen
  \bibfield  {author} {\bibinfo {author} {\bibfnamefont {C.}~\bibnamefont
  {de~Rham}}\ and\ \bibinfo {author} {\bibfnamefont {S.}~\bibnamefont
  {Melville}},\ }\href {https://doi.org/10.1103/PhysRevLett.121.221101}
  {\bibfield  {journal} {\bibinfo  {journal} {Phys. Rev. Lett.}\ }\textbf
  {\bibinfo {volume} {121}},\ \bibinfo {pages} {221101} (\bibinfo {year}
  {2018})},\ \Eprint {https://arxiv.org/abs/1806.09417} {arXiv:1806.09417
  [hep-th]} \BibitemShut {NoStop}%
\bibitem [{\citenamefont {Crisostomi}\ \emph
  {et~al.}(2019{\natexlab{a}})\citenamefont {Crisostomi}, \citenamefont
  {Koyama}, \citenamefont {Langlois}, \citenamefont {Noui},\ and\ \citenamefont
  {Steer}}]{Crisostomi:2018bsp}%
  \BibitemOpen
  \bibfield  {author} {\bibinfo {author} {\bibfnamefont {M.}~\bibnamefont
  {Crisostomi}}, \bibinfo {author} {\bibfnamefont {K.}~\bibnamefont {Koyama}},
  \bibinfo {author} {\bibfnamefont {D.}~\bibnamefont {Langlois}}, \bibinfo
  {author} {\bibfnamefont {K.}~\bibnamefont {Noui}},\ and\ \bibinfo {author}
  {\bibfnamefont {D.~A.}\ \bibnamefont {Steer}},\ }\href
  {https://doi.org/10.1088/1475-7516/2019/01/030} {\bibfield  {journal}
  {\bibinfo  {journal} {JCAP}\ }\textbf {\bibinfo {volume} {01}},\ \bibinfo
  {pages} {030}},\ \Eprint {https://arxiv.org/abs/1810.12070} {arXiv:1810.12070
  [hep-th]} \BibitemShut {NoStop}%
\bibitem [{\citenamefont {Armendariz-Picon}\ \emph {et~al.}(2000)\citenamefont
  {Armendariz-Picon}, \citenamefont {Mukhanov},\ and\ \citenamefont
  {Steinhardt}}]{Armendariz-Picon:2000nqq}%
  \BibitemOpen
  \bibfield  {author} {\bibinfo {author} {\bibfnamefont {C.}~\bibnamefont
  {Armendariz-Picon}}, \bibinfo {author} {\bibfnamefont {V.~F.}\ \bibnamefont
  {Mukhanov}},\ and\ \bibinfo {author} {\bibfnamefont {P.~J.}\ \bibnamefont
  {Steinhardt}},\ }\href {https://doi.org/10.1103/PhysRevLett.85.4438}
  {\bibfield  {journal} {\bibinfo  {journal} {Phys. Rev. Lett.}\ }\textbf
  {\bibinfo {volume} {85}},\ \bibinfo {pages} {4438} (\bibinfo {year}
  {2000})},\ \Eprint {https://arxiv.org/abs/astro-ph/0004134}
  {arXiv:astro-ph/0004134} \BibitemShut {NoStop}%
\bibitem [{\citenamefont {Armendariz-Picon}\ \emph {et~al.}(2001)\citenamefont
  {Armendariz-Picon}, \citenamefont {Mukhanov},\ and\ \citenamefont
  {Steinhardt}}]{Armendariz-Picon:2000ulo}%
  \BibitemOpen
  \bibfield  {author} {\bibinfo {author} {\bibfnamefont {C.}~\bibnamefont
  {Armendariz-Picon}}, \bibinfo {author} {\bibfnamefont {V.~F.}\ \bibnamefont
  {Mukhanov}},\ and\ \bibinfo {author} {\bibfnamefont {P.~J.}\ \bibnamefont
  {Steinhardt}},\ }\href {https://doi.org/10.1103/PhysRevD.63.103510}
  {\bibfield  {journal} {\bibinfo  {journal} {Phys. Rev. D}\ }\textbf {\bibinfo
  {volume} {63}},\ \bibinfo {pages} {103510} (\bibinfo {year} {2001})},\
  \Eprint {https://arxiv.org/abs/astro-ph/0006373} {arXiv:astro-ph/0006373}
  \BibitemShut {NoStop}%
\bibitem [{Pla()}]{Planck_Legacy_Archive}%
  \BibitemOpen
  \href@noop {} {}\bibinfo {howpublished}
  {\url{http://pla.esac.esa.int/pla/}}\BibitemShut {NoStop}%
\bibitem [{\citenamefont {Aghanim}\ \emph
  {et~al.}(2020{\natexlab{b}})\citenamefont {Aghanim} \emph
  {et~al.}}]{Planck:2019nip}%
  \BibitemOpen
  \bibfield  {author} {\bibinfo {author} {\bibfnamefont {N.}~\bibnamefont
  {Aghanim}} \emph {et~al.} (\bibinfo {collaboration} {Planck}),\ }\href
  {https://doi.org/10.1051/0004-6361/201936386} {\bibfield  {journal} {\bibinfo
   {journal} {Astron. Astrophys.}\ }\textbf {\bibinfo {volume} {641}},\
  \bibinfo {pages} {A5} (\bibinfo {year} {2020}{\natexlab{b}})},\ \Eprint
  {https://arxiv.org/abs/1907.12875} {arXiv:1907.12875 [astro-ph.CO]}
  \BibitemShut {NoStop}%
\bibitem [{\citenamefont {Hirano}\ \emph {et~al.}(2019)\citenamefont {Hirano},
  \citenamefont {Kobayashi},\ and\ \citenamefont {Yamauchi}}]{Hirano:2019scf}%
  \BibitemOpen
  \bibfield  {author} {\bibinfo {author} {\bibfnamefont {S.}~\bibnamefont
  {Hirano}}, \bibinfo {author} {\bibfnamefont {T.}~\bibnamefont {Kobayashi}},\
  and\ \bibinfo {author} {\bibfnamefont {D.}~\bibnamefont {Yamauchi}},\ }\href
  {https://doi.org/10.1103/PhysRevD.99.104073} {\bibfield  {journal} {\bibinfo
  {journal} {Phys. Rev. D}\ }\textbf {\bibinfo {volume} {99}},\ \bibinfo
  {pages} {104073} (\bibinfo {year} {2019})},\ \Eprint
  {https://arxiv.org/abs/1903.08399} {arXiv:1903.08399 [gr-qc]} \BibitemShut
  {NoStop}%
\bibitem [{\citenamefont {Crisostomi}\ \emph
  {et~al.}(2019{\natexlab{b}})\citenamefont {Crisostomi}, \citenamefont
  {Lewandowski},\ and\ \citenamefont {Vernizzi}}]{Crisostomi:2019yfo}%
  \BibitemOpen
  \bibfield  {author} {\bibinfo {author} {\bibfnamefont {M.}~\bibnamefont
  {Crisostomi}}, \bibinfo {author} {\bibfnamefont {M.}~\bibnamefont
  {Lewandowski}},\ and\ \bibinfo {author} {\bibfnamefont {F.}~\bibnamefont
  {Vernizzi}},\ }\href {https://doi.org/10.1103/PhysRevD.100.024025} {\bibfield
   {journal} {\bibinfo  {journal} {Phys. Rev. D}\ }\textbf {\bibinfo {volume}
  {100}},\ \bibinfo {pages} {024025} (\bibinfo {year} {2019}{\natexlab{b}})},\
  \Eprint {https://arxiv.org/abs/1903.11591} {arXiv:1903.11591 [gr-qc]}
  \BibitemShut {NoStop}%
\bibitem [{\citenamefont {Sheather}(2004)}]{Sheather:2004}%
  \BibitemOpen
  \bibfield  {author} {\bibinfo {author} {\bibfnamefont {S.~J.}\ \bibnamefont
  {Sheather}},\ }\href {https://doi.org/10.1214/088342304000000297} {\bibfield
  {journal} {\bibinfo  {journal} {Statistical Science}\ }\textbf {\bibinfo
  {volume} {19}},\ \bibinfo {pages} {588 } (\bibinfo {year}
  {2004})}\BibitemShut {NoStop}%
\end{thebibliography}%

\end{document}